\documentclass[english,twocolumn]{revtex4-2}
\usepackage{lmodern}

\usepackage[T1]{fontenc}
\usepackage[latin9]{inputenc}
\usepackage{xcolor}
\usepackage{pdfcolmk}
\usepackage{float}
\usepackage{amsmath}
\usepackage{amssymb}
\usepackage{graphicx}
\PassOptionsToPackage{normalem}{ulem}
\usepackage{ulem}

\makeatletter

\providecolor{lyxadded}{rgb}{0,0,1}
\providecolor{lyxdeleted}{rgb}{1,0,0}

\DeclareRobustCommand{\lyxsout}[1]{\ifx\\#1\else\sout{#1}\fi}

\makeatother

\usepackage{babel}
\begin{document}
\title{Many-species ecological fluctuations as a jump process from the brink
of extinction}
\author{Thibaut Arnoulx de Pirey and Guy Bunin}
\affiliation{Department of Physics, Technion-Israel Institute of Technology, Haifa
32000, Israel}
\begin{abstract}
Highly-diverse ecosystems exhibit a broad distribution of population
sizes and species turnover, where species at high and low abundances
are exchanged over time. We show that these two features generically
emerge in the fluctuating phase of many-variable model ecosystems
with disordered species interactions\textcolor{black}{, when species
are supported by migration from outside the system at a small rate.
We show that these and other phenomena can be understood through the
existence of a scaling regime in the limit of small migration, in
which large fluctuations and long timescales emerge.} We construct
an exact analytical theory for this asymptotic regime, that provides
scaling predictions on timescales and abundance distributions that
are verified exactly in simulations.

In this regime, \textcolor{black}{a clear separation emerges between
rare and abundant species at any given time, despite species moving
back and forth between the rare and abundant subsets. The number of
abundant species is found to lie strictly below a well-known stability
bound, maintaining the system away from marginality. At the same time,
other measures of diversity, which also include some of the rare species,
go above this bound.}

In the asymptotic limit where the migration rate goes to zero, trajectories
of individual species abundances are described by non-Markovian jump-diffusion
processes, which proceeds as follows: A rare species remains so for
some time, then experiences a jump in population sizes after which
it becomes abundant (a species turnover event) and later sees its
population size gradually decreasing again until rare, due to the
competition with other species. The asymmetry of abundance trajectories
under time-reversal is maintained at small but finite migration rate.
These features may serve as fingerprints of endogenous fluctuations
in highly-diverse ecosystems.
\end{abstract}
\maketitle

\section{Introduction\label{sec:Introduction}}

In ecological communities, interactions between species can drive
changes in population sizes. For few-species communities, experiments
find dynamics including stable equilibria, periodic oscillations and
chaos \citep{beninca_Species_2015,fussmann_Crossing_2000,gause_Experimental_1934},
which are explained in terms of dynamical models of interacting populations,
such as Lotka-Volterra or resource-competition models \citep{hofbauer_Evolutionary_1998}.
Yet many natural ecosystems, from microbes in a grain of soil to plants
in a rainforest, can be staggeringly diverse. Nevertheless, the dynamics
of such highly-diverse communities are far less understood.

Observations on highly diverse systems show that the distribution
of species abundances (population sizes) at a given time is often
very broad, with many species at very low population size \citep{grilli_Macroecological_2020,ser2018ubiquitous}.
Time fluctuations in abundances can be very large, with ``blooms''
and significant species turnover, where the species at high and low
abundances are exchanged over time \citep{martin-platero_High_2018,ignacio-espinoza_Longterm_2020,miele2020core}.

Theoretically, dynamical models of interacting populations with many
variables can be notoriously challenging to analyze. They are parameterized
by very many parameters describing the interactions between species,
which are unknown and often unrealistic to obtain from measurements.
This has prompted a change of paradigm (following similar ideas in
physics and other fields), replacing unknown parameters by randomly
sampled ones \citep{gardner1970connectance,may_Will_1972}, and looking
for typical and universal properties of the many-variable systems.
Both physicists and ecologists aim to classify the different broad
behaviors and the robust features of each, formalized within physics
as ``phases''. An important contribution of statistical physics
is the ability to provide mathematical frameworks, giving systematic
answers to questions on these robust properties. This work provides
such a framework for one such challenging phase.

Two distinct phases that have recently attracted much attention, are
a phase where the abundances of different species reach a fixed point,
and another where they fluctuate indefinitely \citep{opper_phase_1992,bunin_Ecological_2017}.
These distinct behaviors have been observed in controlled experiments
where properties of microbial communities are varied \citep{hu_Emergent_2022},
highlighting the power of robust theoretical predictions when applied
to ecological phenomena.

Most of the theoretical work so far has been devoted to the fixed
point phase, and many of its properties are well understood, including
the abundance distribution, and limits on the fixed point stability
that signal the transition to the fluctuating phase \citep{opper_phase_1992,bunin_Ecological_2017,kessler_generalized_2015}.
Predictions obtained theoretically for the fixed-point phase stand
in contrast with empirically observed broad abundance distributions,
and are also unable to account for situations where large abundance
fluctuations are observed. The dynamical phase, which is the focus
of this work, holds promise of addressing these limitations.

Much less has been known about the dynamical phase. For well-mixed
systems (no explicit space), that are coupled to the outside by a
migration of all species, simulations have shown that the system reached
a stationary chaotic state \citep{roy_Numerical_2019}. Similar results
have been obtained from simulations of coupled spatial locations \citep{roy_Complex_2020,pearce_Stabilization_2020}.
However, for well-mixed systems in the absence of migration, a dynamical
slowdown is observed, along with large population fluctuations \citep{roy_Numerical_2019,pearce_Stabilization_2020}.
Analytical results for many-species fluctuating dynamics have been
derived when interactions are fully anti-symmetric, in which case
a stationary chaotic state is reached even without migration \citep{pearce_Stabilization_2020}.
Yet, this state is sensitive to the anti-symmetry that is not expected
to hold generally in nature \citep{pearce_Stabilization_2020}. A
few-species model featuring dynamical slowdown and large fluctuations
is the three-species rock-paper-scissors dynamics, cycling between
and ever-closer to three unstable fixed points (a heteroclinic orbit)
\citep{may_Nonlinear_1975}. This elegant model serves as an instructive
analogy for many-species dynamics \citep{pearce_Stabilization_2020,osullivan_Intrinsic_2021,arnoulxdepirey_Aging_2023},
yet is only of limited relevance to many-species properties such as
diversity, abundance distributions, and stability. A many-species
exactly-solvable dynamical toy model that features dynamical slowdown
was introduced in \citep{arnoulxdepirey_Aging_2023}. Yet due to its
special structure, this model did not include key features of ecological
systems. In particular, questions relating diversity and linear stability
that appear in many other systems cannot be addressed, and it also
did not include migration that interrupts the dynamical slowdown process.

In this work, we provide a systematic analytical framework for the
dynamical phase, for the Lotka-Volterra model with randomly sampled
interactions, in the limit of many species, and when migration rates
are small. In this phase, fluctuations in population sizes are caused
solely by interactions between species, without changes in the environment.
We show that it is precisely in the limit of low migration, where
fluctuations in population sizes get slower and larger, that many
of the striking signatures of this phase emerge. These features, which
are now listed, include commonly observed traits of high-diversity
natural ecosystems, such as a broad abundance distribution and species
turnover, see points 1 and 2 below. In addition, they shed new light
on the definition and measurement of diversity when species turnover
is involved, see point 4, and uncover new phenomena that could be
used as fingerprints of endogenous fluctuations, see point 5.
\begin{enumerate}
\item At any given time, abundances are broadly distributed, with many rare
species with population size close to the minimal value set by the
migration and many species with small population sizes but yet much
larger than the minimal one. We show analytically that over this intermediate
range, the abundance distribution scales as a power law with exponent
$-1$ and characterize the corrections to this power law behavior
when the migration rate is finite. Broad distributions have been measured
in natural ecosystems \citep{locey2016scaling,ser2018ubiquitous},
but their form and origin have been debated \citep{azaele_statistical_2016,grilli_Macroecological_2020}.
In contrast, in the fixed point phase, the number of species in this
intermediate range is small and the abundances are not broadly distributed
\citep{bunin_Ecological_2017}.
\item The dynamics exhibit \emph{species turnover}, where the species at
high abundance are exchanged over time with species that were at low
abundances. In particular, we show that there are always species at
low abundances that are able to grow. This is in contrast to the fixed
point phase, where rare species cannot invade.
\item A long timescale emerges, possibly extending over many generations:
as the migration gets lower, temporal changes in both abundances and
growth rates become slower. This timescale scales as the absolute
value of the logarithm of the migration rate, a prediction that might
be directly tested in controlled experiments. Understanding the various
timescales involved is key to understanding ecological dynamics \citep{hastings2010timescales}
and this work identifies a new collective mechanism through which
long timescales are robustly self-generated by species interactions
in highly-diverse model ecosystems. The combination of species that
can invade and long timescales allows even for species that are very
rare to reach high abundance later. Hence, rare species at a given
time can be important in the future, and the distinction of rare versus
abundant species is not fixed in time. Due to the long timescale,
the species at high abundance lie close to a fixed point, which would
have been stable in the absence of the other species.
\item How do many species coexist in highly-diverse ecosystems is one of
the key questions in ecology. Turnover events and broad abundance
distributions raise questions as to how one might even \emph{define}
diversity. We show that while the list of abundant species changes
in time, the fraction above any given threshold abundance fluctuates
only by a little. Furthermore, in the low migration limit, the number
of species with high and intermediate abundance is well-defined, namely
insensitive to the precise choice of the thresholds. Finally, we show
that the number of high-abundance species, which lie close to a fixed
point, is strictly below the stability bound (known as the ``May
bound'' \citep{may_Will_1972}). Equivalently, this fixed point is
fully stable (as opposed to marginally-stable). In contrast, the number
of high-abundance and intermediate-abundance species together is not
constrained by the stability bound and exceeds it. Therefore measures
of species-richness that also probe the power law region of the species
abundance distribution are not constrained by May's stability bound.
In fact, most of the species whose population sizes are well above
the minimal value set by migration, have small abundances.
\item Lastly, we show that ``bloom'' dynamics, where a species grows from
rare to abundant, until eventually going back to rare, are strongly
asymmetric under time-reversal: The trajectory of the population size
starts with a quick increase and decreases back gradually. This is
an interesting signature of purely endogenous fluctuations in high-diversity
ecosystems that could be looked for in empirical time series.
\end{enumerate}
In addition to the above predictions, we also consider isolated systems
(namely, without migration from the outside) and show that there,
the timescale and size of the abundance fluctuations will continue
to grow in time indefinitely. This will inevitably lead to extinctions
of many species, once the finite size of the populations is taken
into account.

The core of our argument is built on identifying the appropriate transformations
of time and abundances, for which all dynamical properties (including
steady-state distributions and two-time correlations) collapse for
different values of migration, when migration is low. These scaling
relations are verified exactly in collapse of simulation data. We
obtain a well-defined stochastic process for these transformed variables
in the limit of small but positive migration rate. The resulting picture
has features that set it apart from generic many-variable chaotic
dynamics. First, abundances follow non-Markovian jump-diffusion dynamics.
Second, even though the dynamics are slow, the system is maintained
strictly away from marginality, in contrast with glassy systems \citep{cugliandolo_Analytical_1993,kurchan_Phase_1996}.
Last, the emergence of the long timescale results from dynamical slowdown
(aging) that would have continued indefinitely at zero migration.
Yet, this slowdown is not due to the existence of a rough landscape.
We trace all the unique phenomenology back to the possibility of extinctions
in the absence of migration. In other words, this is a consequence
of the multiplicative nature of the dynamics, which are constrained
to positive population sizes with absorbing boundaries at zero population
sizes.

\begin{figure}[!tb]
\begin{centering}
\includegraphics[viewport=0bp 0bp 949bp 1376bp,width=1\columnwidth]{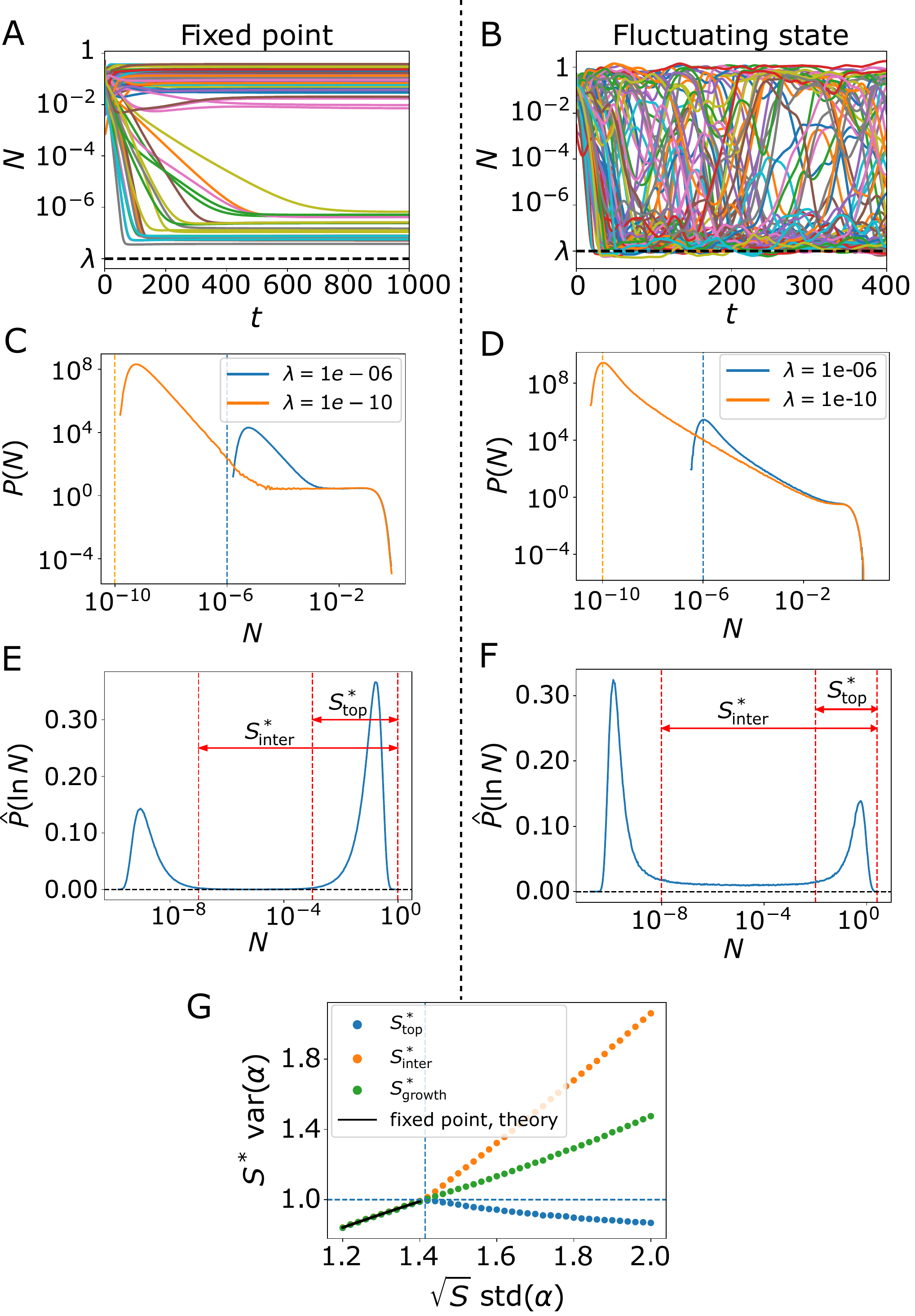}
\par\end{centering}
\caption{\textbf{Species richness and species abundance distributions at fixed
points (left column) and a persistently fluctuating state (right column).}
(A) Dynamics reaching a fixed point. (B) Persistent dynamics, where
the abundances fluctuate indefinitely in the range $\lambda\lesssim N_{i}\lesssim1$.
(C-F) The corresponding distributions for $N$ and $\ln N$, denoted
by $P(N)$ and $\hat{P}(\ln N)$ respectively. The distributions contain
three parts (see E,F): a top part, which remains $O\left(\lambda^{0}\right)$
when $\lambda\to0^{+}$, containing $S_{\text{top}}^{*}$ species;
an intermediate part, at $\lambda\ll N\ll1$; and a low part at $O(\lambda)$.
The intermediate part contains a finite fraction of the species in
the dynamical phase, but not in the fixed point phase.\textcolor{red}{{}
}(G) The three definitions of species richness, normalized by $\text{var}(\alpha)$
so that $1$ is the fixed-point stability bound, as a function of
the interaction variability $\sigma$. For $\sigma<\sigma_{c}$ a
fixed point is reached, all three definitions coincide and lie below
the stability bound, and agree with known theory (solid line). At
$\sigma=\sigma_{c}$ the stability bound is reached. Beyond it, there
are persistent fluctuations, and the three definitions no longer coincide:
$S_{\mathrm{top}}^{*}$ is \emph{lower} than the stability bound,
while $S_{\mathrm{inter}}^{*}$ and $S_{\text{growth}}^{*}$ are above
it. The diversities are obtained by solving the rescaled dynamics
defined in Sec. \ref{sec:Rescaled-dynamics}, and agree quantitatively
with careful analysis of the full equations of motion (see Sec. \ref{sec:robustness}).
Simulation parameters for all figures are given in Appendix \ref{sec:Numerical-methods}.\label{fig:richness-SAD-1}}
\end{figure}

\section{Model definition, the two phases\label{sec:Model-definition}}

We start with the Lotka-Volterra system of equations

\begin{equation}
\dot{N}_{i}=N_{i}\left(1-N_{i}-\sum_{j(\neq i)}\alpha_{ij}N_{j}\right)+\lambda\,,\label{eq:multibody}
\end{equation}
for $i=1\dots S$ where $S$ is the total number of species. The variables
$N_{i}$ represent the abundances (population sizes) of the different
species, and so $N_{i}\ge0$ at the initial time and is guaranteed
to remain so throughout. $\lambda\ge0$ represents migration from
an external source, which for simplicity is taken to be the same for
all $i$.

Eq. (\ref{eq:multibody}) is a standard rescaling \citep{may2007theoretical,takeuchi1996global,bunin_Ecological_2017,barbier_Generic_2018}
of the Lotka-Volterra equation $\dot{n}_{i}=r_{i}n_{i}/k_{i}(k_{i}-\sum_{j}A_{ij}n_{j})+D_{i}$.
For simplicity, we take all $r_{i}=r$. Eq. (\ref{eq:multibody})
is obtained by rescaling time by $r$, and introducing the non-dimensional
parameters $N_{i}\equiv n_{i}/k_{i}$, $\alpha_{ij}\equiv A_{ij}k_{j}/k_{i}$
and $\lambda\equiv D_{i}/\left(rk_{i}\right)$. Below we study the
limit $\lambda\ll1$, but assume that absolute population sizes $n_{i}$,
whose minimal size is around $D_{i}/r$, are always large enough so
that demographic stochasticity can be neglected. The large parameter
$\left|\ln\lambda\right|$ will play an important role in the following,
and since the absolute population sizes $k_{i}$ can be large (e.g.,
up to $10^{10}$ bacterial cells in one milliliter), $\left|\ln\lambda\right|$
can be reasonably large in ecologically relevant settings.

The limit of large $S$ is relevant to many natural high-diversity
ecosystems, with hundreds to thousands of species, in communities
from microbes to trees \citep{wright2002plant,grilli_Macroecological_2020}.
We assume that $\mathbb{\alpha}$ is a random matrix with Gaussian
entries, and for mathematical convenience we carry most of the analysis
in the case where the interaction coefficients are all sampled independently
from each other such that $\langle\alpha_{ij}\rangle=\mu/S$ and $\langle\alpha_{ij}\alpha_{kl}\rangle-\langle\alpha_{ij}\rangle\langle\alpha_{kl}\rangle=\sigma^{2}\delta_{ik}\delta_{kl}/S$.
By using combinations of numerical simulations and analytical calculations,
we later show that the qualitative picture presented in this paper
is nonetheless robust to the addition of correlations (positive or
negative) of the interaction coefficients within pairs of species,
see Sec. \ref{sec:robustness} .

The system exhibits different dynamical behaviors depending on the
parameters \citep{bunin_Ecological_2017}. When $\mu>0$ and the heterogeneity
of interactions is smaller than a critical value $\sigma<\sigma_{c}$
the dynamics reach a fixed point, see Fig. \ref{fig:richness-SAD-1}(A).
In it, some species are absent and others remain present. This definition
is straightforward when $\lambda=0$, where a fixed point $dN_{i}/dt=0$
in Eq. (\ref{eq:multibody}) implies either $N_{i}=0$ that are the
absent species, or $N_{i}>0$ which are the present species. At the
stable fixed point reached, the extinct species have negative growth
rates $\dot{N}_{i}/N_{i}<0$ and so cannot invade, and the subset
of present species is linearly stable. For small $\lambda>0$, the
absent species are now at values $N_{i}$ of order $\lambda$ and
the present species are unaffected by the small $\lambda$. The requirements
for a stable fixed point remain the same.

Above $\sigma>\sigma_{c}$, the system evolves indefinitely, see Fig.
\ref{fig:richness-SAD-1}(B), without ever settling at a stable fixed
point (in fact, such stable, uninvadable fixed points do not exist
\citep{ros_Generalized_2022}).

\section{Phenomenology\label{sec:Phenomenology}}

We now describe key features of the dynamics when $\sigma\geq\sigma_{c}$
and $\lambda\ll1$, starting with the Species Abundance Distribution
(SAD), and then turn to the dynamics.

\subsection{Species abundance and stability\label{subsec:Species-abundance-and}}

The species abundance distribution $P(N)$ in the dynamically-fluctuating
phase is shown in Fig. \ref{fig:richness-SAD-1}(D,F). It spans many
orders of magnitude, ranging from $O(1)$ population sizes to order
$O(\lambda)$ populations sizes. This is indeed the order of magnitude
of the minimal value allowed by migration, which we call the migration
floor in the following. Compared to the equilibrium situation, Fig.
\ref{fig:richness-SAD-1}(C,E), there are many more species with abundances
in the intermediate range $\lambda\ll N\ll1$ and the fraction of
species there remains finite even for small $\lambda$, see Fig. \ref{fig:richness-SAD-1}(F).
In this range, $P(N)$ appears to be approximately a power law. The
value of $P(\ln N)$ changes slowly, so the distribution of the abundances,
$P(N)$, is expected to behave roughly as $N^{-1}$ \citep{pearce_Stabilization_2020,dalmedigos_Dynamical_2020},
see Fig. \ref{fig:richness-SAD-1}(C). In Sec. \ref{subsec:Species-Abundance-Distribution},
we show that in the $\lambda\to0^{+}$ limit the power law is indeed
exactly $N^{-1}$, and refine this picture with precise corrections
to this $N^{-1}$ behavior.

A common way to define the number of present species (the ``species
richness'') is by those whose abundance lies above some value. We
consider two definitions based on this idea, and an additional criterion
based on the invasion growth rate. The three proposed definitions
of species richness are as follows:
\begin{enumerate}
\item $S_{\text{top}}^{*}$ is the number of species belonging to the right
peak in Fig. \ref{fig:richness-SAD-1}(F). Their abundance $N_{i}$
remains finite even for small $\lambda$. These are the top or abundant
species.
\item $S_{\text{inter}}^{*}$ includes the number of ``intermediate''
species. It is all species except those belonging to the left peak
in Fig. \ref{fig:richness-SAD-1}(F). They satisfy $N_{i}\gg\lambda$.
\item A third definition, $S_{\text{growth}}^{*}$, can be obtained through
invasion experiments. If the population size of species $i$ is set
to a value that is small but well-above the migration floor, $\lambda\ll N_{i}^{\mathrm{new}}\ll1$,
and keeping all other $N_{j}$ unchanged, species $i$ will grow with
$\left[\dot{N}_{i}/N_{i}\right]_{\text{invasion}}=g_{i}>0$. $g_{i}$
is called the ``invasion growth rate'' (see, e.g. \citep{marrow_coevolution_1992,hofbauer_Permanence_2022}).
For the dynamics Eq. (\ref{eq:multibody}), $g_{i}=1-\sum_{j(\neq i)}\alpha_{ij}N_{j}$.
$S_{\text{growth}}^{*}$ counts the number of species with positive
growth rate.
\end{enumerate}
At the fixed point phase, all three definitions coincide, $S_{\text{top}}^{*}=S_{\text{inter}}^{*}=S_{\text{growth}}^{*}=S^{*}$,
when $\lambda$ is small. In the fluctuating phase, given the presence
of species in the range $\lambda\ll N\ll1$, one might worry that
species richness $S_{\text{top}}^{*}$ is not well-defined, in that
it relies on an arbitrary cutoff on the abundances to decide which
species are ``present'' or ``absent'', with a similar concern
for $S_{\text{inter}}^{*}$. In the limit $\lambda\to0^{+}$, we show
below that the peaks in see Fig. \ref{fig:richness-SAD-1}(F) become
narrow compared to $\left|\ln\lambda\right|$, and therefore the definitions
for $S_{\text{top}}^{*},S_{\text{inter}}^{*}$ become sharp. Focusing
on $S_{\text{top}}^{*}$, we argue in Sec. \ref{sec:robustness} that
for $\lambda$ small but reasonable for ecological applications, these
quantities can be measured in practice, and are not far from their
asymptotic values.

The values of $S_{\text{top}}^{*},S_{\text{inter}}^{*},S_{\text{growth}}^{*}$
are plotted in Fig. \ref{fig:richness-SAD-1}(G). They are compared
to a bound on species richness coming from linear stability (horizontal
dashed line): If a subset of $S^{*}$ abundant species is at a fixed
point, meaning that the abundances satisfy $1-N_{i}-\sum_{j(\neq i)}\alpha_{ij}N_{j}=0$,
this fixed point will typically be linearly stable (ignoring other
species) if $S^{*}\mathrm{var}(\alpha)\leq1$ \citep{may_Will_1972,bunin_Ecological_2017}.
In the fixed point phase, it is known that $S^{*}\mathrm{var}(\alpha)<1$
up to the transition \citep{bunin_Ecological_2017}; thus, the stability
bound is not saturated in this phase. As shown in Fig. \ref{fig:richness-SAD-1}(G),
after crossing the transition, we have again $S_{\text{top}}^{*}\mathrm{var}(\alpha)<1$,
meaning that the richness of abundant species lies\emph{ strictly
below the stability bound}. In Sec. \ref{subsec:Diversities-revisited},
we will further show that the subset of abundant species lies at any
time in the vicinity of a fixed point, which is thus linearly stable.
This is perhaps surprising, compared to the symmetric case where the
stability bound is saturated \citep{biroli_Marginally_2018}, and
given that the dynamics are slow, as shown below. We return to this
result in Sec. \ref{subsec:Diversities-revisited}.

In contrast, $S_{\text{inter}}^{*}\mathrm{var}(\alpha)$ and $S_{\text{growth}}^{*}\mathrm{var}(\alpha)$
continue to grow above the bound. This is not in contradiction to
the stability bound, since $S_{\text{inter}}^{*}$ and $S_{\text{growth}}^{*}$
count species with intermediate abundances $\lambda\ll N_{i}\ll1$,
and some of them have positive growth rates, so do not satisfy the
condition $\dot{N}_{i}/N_{i}\simeq1-N_{i}-\sum_{j(\neq i)}\alpha_{ij}N_{j}\simeq0$.
The proportion of species with intermediate abundances ($\lambda\ll N\ll1$)
among those with population size well above the migration floor ($N\gg\lambda$)
grows with the standard deviation of the interactions ${\rm std}(\alpha)$,
as can be seen in Fig. \ref{fig:richness-SAD-1}(G).

\subsection{Dynamics and timescales\label{subsec:Dynamics-and-timescales}}

\begin{figure}[!tb]
\begin{centering}
\includegraphics[width=1\columnwidth]{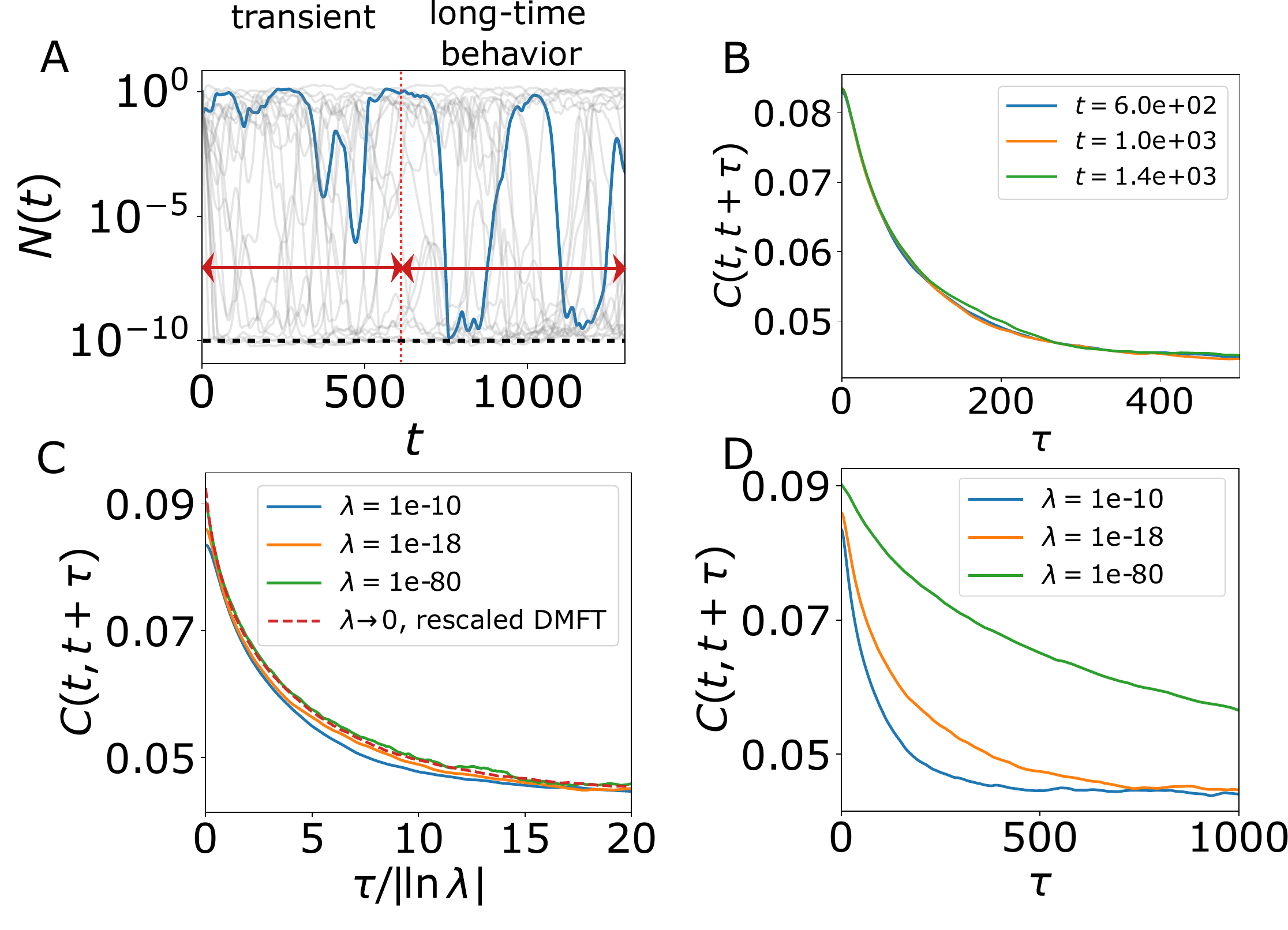}
\par\end{centering}
\caption{\textbf{Species dynamics and correlations.} (A) Species are initialized
with order-one values. We follow $\ln N_{i}(t)$. In the transient
regime, the changes in $\ln N_{i}(t)$ are comprised of three elements:
downward motion, roughly in straight line (maintaining its slope for
a large part of the decay between largest and smallest values, corresponding
to exponential decay of $N_{i}$); upward motion, roughly in a straight
line; and slow changes at high population values, $N_{i}(t)\simeq O(1)$.
The dynamics slow down, with the time spent in each of these elements
increasing with the time since the start of the dynamics, see Sec.
\ref{sec:Phenomenology:-Epoch-I}. As the excursions grow longer,
the average value of $\ln N_{i}(t)$ decreases linearly in $t$. This
transient regime ends when a finite fraction of species reach the
migration floor $N_{i}(t)\simeq O(\lambda)$, after a time of order
$\left|\ln\lambda\right|$. At long times, after the transient regime
is over, $\ln N_{i}(t)$ performs the three dynamical elements described
above, and a forth one, slowly changing around the migration floor.
The vertical red dashed line marks the crossover between these two
regimes. (B) At long times, timescales do not grow in time anymore,
as seen in the correlation function that depends only on time differences,
$C(t,t')=C(t-t')$. (C, D) This timescale is proportional to $\left|\ln\lambda\right|$.
This is demonstrated by the data collapse when plotting $C$ against
$t/\left|\ln\lambda\right|$ presented in panel(C), compared with
the data without rescaling shown in (D).\label{fig:dynamics-1}}
\end{figure}
We now describe the long-time phenomenology of the dynamics, when
$\lambda>0$. The transient regime, see Fig. \ref{fig:dynamics-1}(A),
is later discussed in Sec. \ref{sec:Phenomenology:-Epoch-I} together
with the closely related dynamics of isolated systems, for which $\lambda=0$.
When $\lambda>0$ and at long times, the species abundances fluctuate
forever and their autocorrelation function $C(t,t')\equiv\sum_{i}N_{i}(t)N_{i}(t')/S$
becomes time-translation invariant, namely $C(t,t')=C(t-t')$, see
Fig. \ref{fig:dynamics-1}(B). Crucially, the dynamics become slow
when $\lambda\ll1$ and feature an emergent statistical invariance
between realizations at different $\lambda$ under the rescaling of
time $t\to t/|\ln\lambda|$. As we show in Fig. \ref{fig:dynamics-1}(C),
the autocorrelation functions for different values of $\lambda$,
but identical values of $\sigma$ and $\mu$, indeed collapse to a
single master curve when plotted against $t/\left|\ln\lambda\right|$,
namely $C_{\lambda}(t,t+\vert\ln\lambda\vert s)\to\hat{C}(s)$ when
$\lambda\to0^{+}$. Thus, a unique timescale $\tau\sim|\ln\lambda|$
characterizes the autocorrelation function. We defer the proof of
this result to Sec. \ref{sec:Rescaled-dynamics} where we derive the
rescaled dynamics, that are also solved numerically to obtain the
function $\hat{C}(s)$, see Eqs. (\ref{eq:DMFT2},\ref{eq:dmft-mean2},\ref{eq:dmft-var2}).

Timescales of order $|\ln\lambda|$ are expected to appear, since
it takes a time $t\sim|\ln\lambda|$ for a population to grow from
the migration floor $\lambda$ to $O(1)$ population size under exponential
growth with finite growth rate. Importantly, we find that there is
no shorter timescale $\tau\ll|\ln\lambda|$ relevant for describing
the effective dynamics of a single species. Indeed, the master curve
$\hat{C}(s)$ is regular at $s\to0^{+}$. Recall that $\lambda$ is
non-dimensional, and that $|\ln\lambda|$ can be quite large in ecologically
relevant contexts, see Section \ref{sec:Model-definition}, in which
case fluctuations become correlated over many generation times.

\section{Rescaled dynamics\label{sec:Rescaled-dynamics}}

\begin{figure}[!tb]
\begin{centering}
\includegraphics[width=0.8\columnwidth]{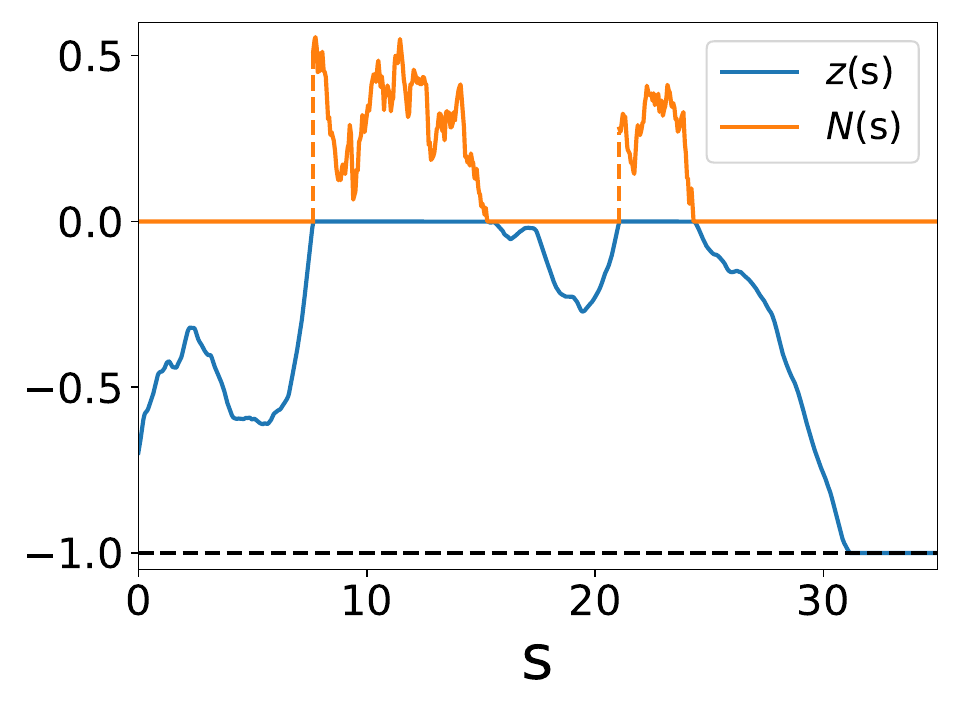}
\par\end{centering}
\caption{\textbf{The rescaled dynamics.} Trajectories of $N$ and $z\equiv\ln(N)/\left|\ln\lambda\right|$,
as a function of the rescaled time $s\equiv t/\left|\ln\lambda\right|$,
in an example run of the limiting rescaled dynamics when $\lambda\to0^{+}$.\textcolor{red}{{}
}\label{fig:The-rescaled-dynamics.}}
\end{figure}

In Sec. \ref{sec:Phenomenology} we described the phenomenology of
the long-time fluctuating dynamics when $0<\lambda\ll1$. We have
seen that the species abundances $N_{i}(t)$ fluctuate over a long
timescale of order $|\ln\lambda|$. Furthermore, the log-abundances
$\ln N_{i}$ dynamically explore values from $O(\ln\lambda)$ to $O(1)$.
These observations motivate defining rescaled variables $s\equiv t/|\ln\lambda|$
and $z_{i}\equiv\ln\left(N_{i}\right)/\left|\ln\lambda\right|$. $z_{i}(s)$
turns out to follow a well-defined stochastic process when $\lambda\to0^{+}$
(that no longer includes any $\lambda$ dependence). It is a single-variable
process describing the probability of trajectories of a single species
within the large system. Subsection \ref{subsec:derivation} is devoted
to the mathematical derivation of this process. Subsection \ref{subsec:description}
describes its properties, and can be read independently of subsection
\ref{subsec:derivation}. Later sections discuss its implications
to the species abundance distribution and species diversity.

\subsection{Derivation\label{subsec:derivation}}

Our starting point is the effective single-species stochastic dynamics
for $N_{i}(t)$, previously obtained in the limit where the number
of species is very large and for any value of $\lambda$ \citep{roy_Numerical_2019}.
For the sake of completeness, we briefly outline the steps leading
to these dynamics. In Eq. (\ref{eq:multibody}), the dynamics of $N_{i}(t)$
are driven by the influence of all other species through $g_{i}(t)=1-\sum_{j}\alpha_{ij}N_{j}(t)$,
which involves the sum of many weakly-correlated contributions. Dynamical
Mean-Field Theory (DMFT), valid when $S\to\infty$, shows that the
$g_{i}(t)$ are identically distributed Gaussian processes, and are
independent for $i\ne j$. This implies that population sizes $N_{i}(t)$
behave as independent realizations of the single-variable stochastic
process,

\begin{equation}
\dot{N}(t)=N(t)(g(t)-N(t))+\lambda\,.\label{eq:DMFT1}
\end{equation}
where the subscript $i$ has been dropped. The first two moments of
the Gaussian process $g(t)$ obey self-consistent closure relations,
that relate the input noise $g(t)$ to the output $N(t)$,
\begin{align}
\left\langle g(t)\right\rangle  & =1-\mu\langle N(t)\rangle\,,\label{eq:dmft_mean1}\\
\langle g(t)g(t')\rangle-\langle g(t')\rangle\langle g(t)\rangle & =\sigma^{2}\langle N(t)N(t')\rangle\,.\label{eq:dmft-var1}
\end{align}
Here the angular brackets $\langle.\rangle$ denote an average over
the realizations of $g(t)$ (and the initial conditions $N(0)$ which
are irrelevant at large times). The derivation of Eq. (\ref{eq:DMFT1})
follows a standard procedure \citep{liu2021dynamics,sompolinsky_Relaxational_1982,mezard_spin_1987,agoritsas_Outofequilibrium_2018}
which was applied to the Lotka-Volterra equations in \citep{roy_Numerical_2019}.
In the long-time limit where $\langle N(t)\rangle\to\langle N\rangle,$
the entire dynamics is controlled by the two-time correlation $C(t,t')\equiv\langle N(t)N(t')\rangle$,
since the Gaussian process $g(t)$ is completely characterized by
its correlations and mean, given in (\ref{eq:dmft_mean1},\ref{eq:dmft-var1}).

In order to study the behavior of these equations when $\lambda\ll1$,
we introduce $z\equiv\ln(N)/\vert\ln\lambda\vert$ and $s\equiv t/\vert\ln\lambda\vert$
so that Eq. (\ref{eq:DMFT1}) becomes

\[
z'(s)=g(s)+\exp(-\vert\ln\lambda\vert(z(s)+1))-\exp(\vert\ln\lambda\vert z(s))\,.
\]
The non-linear terms become impenetrable boundaries when $\lambda\to0^{+}$,
since

\[
\lim_{\lambda\to0^{+}}\exp(-\vert\ln\lambda\vert(z+1))=\begin{cases}
0 & \text{if \ensuremath{z>-1\,,}}\\
+\infty & \text{if \ensuremath{z<-1}\,,}
\end{cases}
\]
and 
\[
\lim_{\lambda\to0^{+}}\exp(\vert\ln\lambda\vert z)=\begin{cases}
0 & \text{if \ensuremath{z<0\,,}}\\
+\infty & \text{if \ensuremath{z>0}\,.}
\end{cases}
\]
Hence the process $z(s)$ is confined between $-1$ and $0$ when
$\lambda\to0^{+}$. The confinement originates from the migration
term and the self-regulation term, proportional to $N_{i}^{2}$, in
Eq. (\ref{eq:multibody}). The effective noise $g(s)$ is not able
to push $z(s)$ to outside of the confining region, because its mean
and variance are finite provided population sizes do not blow up,
see Eqs. (\ref{eq:dmft_mean1},\ref{eq:dmft-var1}). Thus $z(s)$
obeys

\begin{equation}
z'(s)=g(s)+W(z+1)-W(z)\,,\label{eq:DMFT2}
\end{equation}
in the low migration limit, where $W(z)$ and $W(z+1)$ account for
the confining boundaries at $z=0$ and $z=-1$. The autocorrelation
function of $g(s)$ is proportional to the master function $\hat{C}(s)$
introduced in Sec. \ref{sec:Phenomenology}, since $\langle g(s)g(s')\rangle-\langle g(s')\rangle\langle g(s)\rangle=\sigma^{2}\langle N(s)N(s')\rangle=\sigma^{2}\hat{C}(s-s')$.

We now derive the evolution of the abundance $N(s)$. Beyond the fact
that $N(s)$ is the main quantity of interest, such an evolution is
also necessary to derive the self-consistent equations (\ref{eq:dmft_mean1},\ref{eq:dmft-var1})
in the the $\lambda\to0^{+}$ limit, and obtain a closure of the DMFT
equations in terms of the process $z(s)$. When $z(s)<0$ it is clear
from the definition $N(s)=\exp\left(\left|\ln\lambda\right|z(s)\right)$
that $N(s)=0$ in that limit. However, this relation appears ambiguous
in the double limit $z\to0$ and $\lambda\to0^{+}$. To remove the
ambiguity we use the impenetrability condition at the $z=0$ boundary,
namely $W(z(s))=g(s)$ when $z(s)=0$. Since by definition of $z$
and Eq. (\ref{eq:DMFT2}), $W(z(s))=N(s)$, we obtain the relation

\begin{equation}
N(s)=g(s)\Theta(z(s))\,,\label{eq:relaN}
\end{equation}
where $\Theta(z)$ is the Heaviside function with the convention $\Theta(0)=1$.
The existence of this impenetrability condition rests on the facts
that (i) $g(s)$ does not have a white noise component, which follows
from the fact that its mean and variance remain finite when $\lambda\to0^{\text{+}}$
and (ii) that $\hat{C}(s)$ has a well-defined limit when $\lambda\to0^{+}$,
which agrees with numerical simulations (see Fig. \ref{fig:dynamics-1}(C)).
That $\hat{C}(s)$ is well-behaved when $\lambda\to0^{+}$ is for
now assumed and is self-consistently verified is the following Eqs.
(\ref{eq:dmft-mean2},\ref{eq:dmft-var2}). The interpretation of
Eq. (\ref{eq:relaN}) is discussed below in Sec. \ref{subsec:description}.

Using Eq. (\ref{eq:relaN}), the $\lambda\to0^{+}$ limit of the closure
equations (\ref{eq:dmft_mean1},\ref{eq:dmft-var1}) reads

\begin{align}
\langle g(s)\rangle & =1-\mu\langle g(s)\Theta(z(s))\rangle\,,\label{eq:dmft-mean2}\\
\langle g(s)g(s')\rangle-\langle g(s')\rangle\langle g(s)\rangle & =\sigma\langle g(s)g(s')\Theta(z(s))\Theta(z(s'))\rangle\,.\label{eq:dmft-var2}
\end{align}
Importantly, Eqs. (\ref{eq:DMFT2},\ref{eq:dmft-mean2},\ref{eq:dmft-var2})
are independent of $\lambda$. It therefore follows that indeed, $g(s)$
has well-defined correlations in the limit $\lambda\to0^{+}$.

\subsection{Properties of the limit dynamics\label{subsec:description}}

Equation (\ref{eq:DMFT2}) describes the effective evolution of $z_{i}(s)$
for any single species within the many-variable system. It is driven
by a Gaussian noise $g_{i}(s)$ and confined to values $-1\leq z_{i}\leq0$.
The mean and variance of the Gaussian noise $g_{i}(s)$ are determined
self-consistently through Eqs. (\ref{eq:dmft_mean1},\ref{eq:dmft-var1}),
which become Eqs. (\ref{eq:dmft-mean2},\ref{eq:dmft-var2}) in the
limit $\lambda\to0^{+}$. The noise $g_{i}(s)$ can be interpreted
within the original many-species species dynamics Eq. (\ref{eq:multibody})
as the effective growth rate set by all the other species, $g_{i}(s)=1-\sum_{j}\alpha_{ij}N_{j}(s)$,
which can indeed be shown to be Gaussian distributed in the limit
of many species. When $-1<z_{i}(s)<0$, meaning $\lambda\ll N_{i}\ll1$,
the dynamics read $z_{i}'(s)=\dot{N_{i}}(t)/N_{i}(t)=g_{i}(s)$, resulting
in exponential growth or decay of the population sizes under the effective
growth rate set by all the other species.

In the limit $\lambda\to0^{+}$, the dynamics of the population size
$N_{i}(s)$ is related to that of $z_{i}(s)$ by Eq. (\ref{eq:relaN}).
When $z_{i}(s)<0$, Eq. (\ref{eq:relaN}) yields $N_{i}(s)=0$ which
naturally follows from taking the limit $\lambda\to0^{+}$ in the
definition $N_{i}=\exp(|\ln\lambda|z_{i})$. Yet Eq. (\ref{eq:relaN})
goes further, relating $N_{i}$ and $z_{i}$ in the double limit where
both $\lambda\to0^{+}$ and $z_{i}=0$. The interpretation of Eq.
(\ref{eq:relaN}) is that the species with $z_{i}=0$, meaning with
a finite population size $N_{i}$, are in a slowly-changing fixed
point. Indeed, Eq. (\ref{eq:relaN}) can also be obtained from the
many-body dynamics using the slowness of the dynamics discussed in
Sec. \ref{sec:Phenomenology} when $\lambda\to0^{+}$: by Eq. (\ref{eq:multibody}),
slow changes $\dot{N}_{i}(t)\simeq0$ require $N_{i}(t)=g_{i}(t)$
when $N_{i}(t)$ is finite.

Equations (\ref{eq:dmft_mean1},\ref{eq:dmft-var1}) guarantee that
$g(s)$ has finite variance and mean, so that $g(s)$ can not contain
a white noise component. Also, by solving these rescaled DMFT equations
following the procedure detailed in Appendix \ref{sec:Numerical-methods},
we find that $\hat{C}'(0^{+})$ is finite, thus confirming the absence
of fast-time scale in the Lotka-Volterra dynamics, see Fig. \ref{fig:der_c}.
We further show that $\hat{C}'(0^{+})\neq0$, meaning that $g(s)$
is rough, namely nowhere differentiable (like in Brownian motion \footnote{For any process $\eta(s)$ with time-translation invariant correlation
$C(s)$, we have $2|C'(0^{+})|=\lim_{{\rm d}s\to0}\left\langle [\eta(s+{\rm d}s)-\eta(s)]^{2}\right\rangle /{\rm d}s\,$.
Therefore $C'(0^{+})\neq0$ is possible only if the increment $\eta(s+{\rm d}s)-\eta(s)$
scales as $O(\sqrt{{\rm d}s})$ as for Brownian motion.}). The fact that $\hat{C}'(0^{+})\neq0$ follows from Eq. (\ref{eq:fluct_div})
below, see the discussion there. An example run of the rescaled dynamics
is shown in Fig. \ref{fig:The-rescaled-dynamics.}. As can be seen,
$-1\le z\le0$, and $z$ spends finite time intervals at the boundaries,
which is a consequence of the finite memory of the noise $g(s)$.
Indeed, if $g>0$ and $z=0$ at a given time, it will remain so for
a finite amount of time.

Lastly, we discuss important features of the dynamics of the population
sizes $N(s)$ seen in Fig. \ref{fig:The-rescaled-dynamics.}. First,
$N(s)$ is rough since $N=g$ when $z=0$, while $z(s)$ is more smooth
since it is an time integral of $g(s)$. Second, the limiting process
$N(s)$ is not continuous in time and features jumps from $0$ to
a finite value, after which $N(s)$ continuously reaches $0$. The
jumps represent species that grow from very small abundances at a
finite growth rate, and thus their time to reach $N=g$ from a small
fixed value (that doesn't depend on $\lambda$, say $N=10^{-5}$)
is finite in time $t$, and so vanishes in the rescaled time $s$.
These jumps are precisely species turnover events, that drive the
change in the composition of the abundant species.

\section{Phenomenology revisited\label{sec:Phenomenology-revisited}}

\subsection{Diversities revisited\label{subsec:Diversities-revisited}}

In Sec. \ref{subsec:Species-abundance-and}, we proposed three definitions
for diversity $S_{\text{top}}^{*},S_{\text{inter}}^{*},S_{\text{growth}}^{*}$.
Denote $\phi_{\text{top}}\equiv S_{\text{top}}^{*}/S$ and similarly
for $\phi_{\text{inter}},\phi_{\text{growth}}$. All these quantities
are well-defined in the limit $\lambda\to0^{+}$. Indeed, they take
simple forms in terms of the limiting process described in Sec. \ref{sec:Rescaled-dynamics}:
\begin{align}
\phi_{\text{top}} & =\text{Prob}[z=0]\,,\nonumber \\
\phi_{\text{inter}} & =1-\text{Prob}[z=-1]\,,\nonumber \\
\phi_{\text{growth}} & =\text{Prob}[g>0]\,.\label{eq:phis_from_rescaled}
\end{align}

We now return to the discussion of $\phi_{\text{top}}$ and its relation
to the stability bound, see Sec. \ref{subsec:Species-abundance-and}.
As followed from the rescaled dynamics, the abundant species counted
in $\phi_{\text{top}}$ are approximately at a fixed point, while
all other species have negligible abundances, and so do not affect
this fixed point. Thus, one indeed expects the bound $S_{\text{top}}^{*}\text{var}(\alpha)=\sigma^{2}\phi_{\text{top}}\le1$
to hold, as is clear in Fig. \ref{fig:richness-SAD-1}(G). Indeed,
a fixed point with $S_{\text{top}}^{*}$ coexisting species and $S_{\text{top}}^{*}\text{var}(\alpha)>1$
would be typically linearly unstable to perturbations \citep{may_Will_1972,opper_phase_1992,biroli_Marginally_2018}.

A natural question is: Is the stability bound saturated, i.e. $\sigma^{2}\phi_{\text{top}}=1$,
resulting in fixed points of abundant species that are near marginal
stability? One could perhaps expect marginal stability, as the rate
at which low-abundance species are added to the subset of abundant
ones is slow, ``gently'' perturbing the fixed points, and so would
perhaps allow $S_{\text{top}}^{*}$ to increase up to the stability
bound. In addition, marginality is reached in Lotka-Volterra dynamics
\citep{biroli_Marginally_2018}, Eq. (\ref{eq:multibody}), with symmetric
interactions ($\alpha_{ij}=\alpha_{ji}$). And more generally, slow
dynamics are in many cases associated with marginality (see, e.g.,
\citep{muller_Marginal_2015}).

Yet, as was shown in Sec. \ref{subsec:Species-abundance-and}, see
Fig. \ref{fig:richness-SAD-1}(G), $\sigma^{2}\phi_{\text{top}}<1$
so the stability bound is not saturated. Consequently, as $\lambda\to0^{+}$,
the high-abundance species lie at any time in the vicinity of a fixed
point that is linearly stable to perturbations applied to those species.
Note that this fixed point changes over time, as it is destabilized
by the growth of species from rare.

To obtain the bound $\sigma^{2}\phi_{\text{top}}<1$, we prove that
while species turnover is a slow process, the jumps when going from
rare to abundant, seen clearly in Fig. \ref{fig:The-rescaled-dynamics.},
are sufficient to significantly perturb the subset of abundant species
and prevent it from reaching marginal stability. We prove that by
deriving an exact relation that links diversity with temporal fluctuations,
defined as follows. Let $N_{\text{jump}}$ be the size of these jumps.
Let $G$ be the rate of incoming species weighted by $N_{\text{jump}}^{2}$:
that is, the sum of $N_{\text{jump}}^{2}$ over all jumps taking place
in a unit of rescaled time $s$ in the many-species dynamics, and
divided by the number of species $S$. The relation, derived in Appendix
\ref{subsec:Diversity-limited-by}, reads:
\begin{equation}
G=2\left|\hat{C}'\left(0^{+}\right)\right|\left(1-\sigma^{2}\phi_{{\rm top}}\right)\ .\label{eq:fluct_div}
\end{equation}
Here $\hat{C}\left(s\right)$ is the autocorrelation of $N(s)$ in
the rescaled time defined in Sec. \ref{sec:Phenomenology}. $\hat{C}'\left(0^{+}\right)$
is finite, see Fig. \ref{fig:dynamics-1}(C) and Fig. \ref{fig:der_c}
in Appendix \ref{subsec:regularity}. Eq. (\ref{eq:fluct_div}) then
limits $\phi_{\text{top}}$ to be below the stability bound; indeed,
marginal stability $1=\sigma^{2}\phi_{{\rm top}}$ is only possible
if $G=0$, namely species do not perform jumps, in contradiction with
the dynamics in Sec. \ref{sec:Rescaled-dynamics}. Additionally, we
note that $G>0$ implies $\left|\hat{C}'\left(0^{+}\right)\right|>0$,
thus proving that the trajectories $N_{i}(s)$ are rough when $N_{i}(s)\neq0$.

Put differently, the introduction of one new species leads to the
removal of others, with the average number of removed species growing
as one approaches the marginal diversity. The balance, in which one
species is removed for each one introduced, sets $S_{\text{top}}^{*}$,
that only reaches some fraction of the bound $1/\mathrm{var}(\alpha)$.
In dynamics that reach a fixed point, the requirement that all species
involved have positive abundance is known as feasibility \citep{roberts_stability_1974},
and it is what limits the diversity in the fixed point phase, $\sigma<\sigma_{c}$
in Fig. \ref{fig:dynamics-1}(F). Eq. \ref{eq:fluct_div} can be thought
of as an extension of the requirement to dynamics, a form of ``dynamical
feasibility''.

\subsection{Species Abundance Distribution revisited\label{subsec:Species-Abundance-Distribution}}

\begin{figure}[!tb]
\begin{centering}
\includegraphics[width=1\columnwidth]{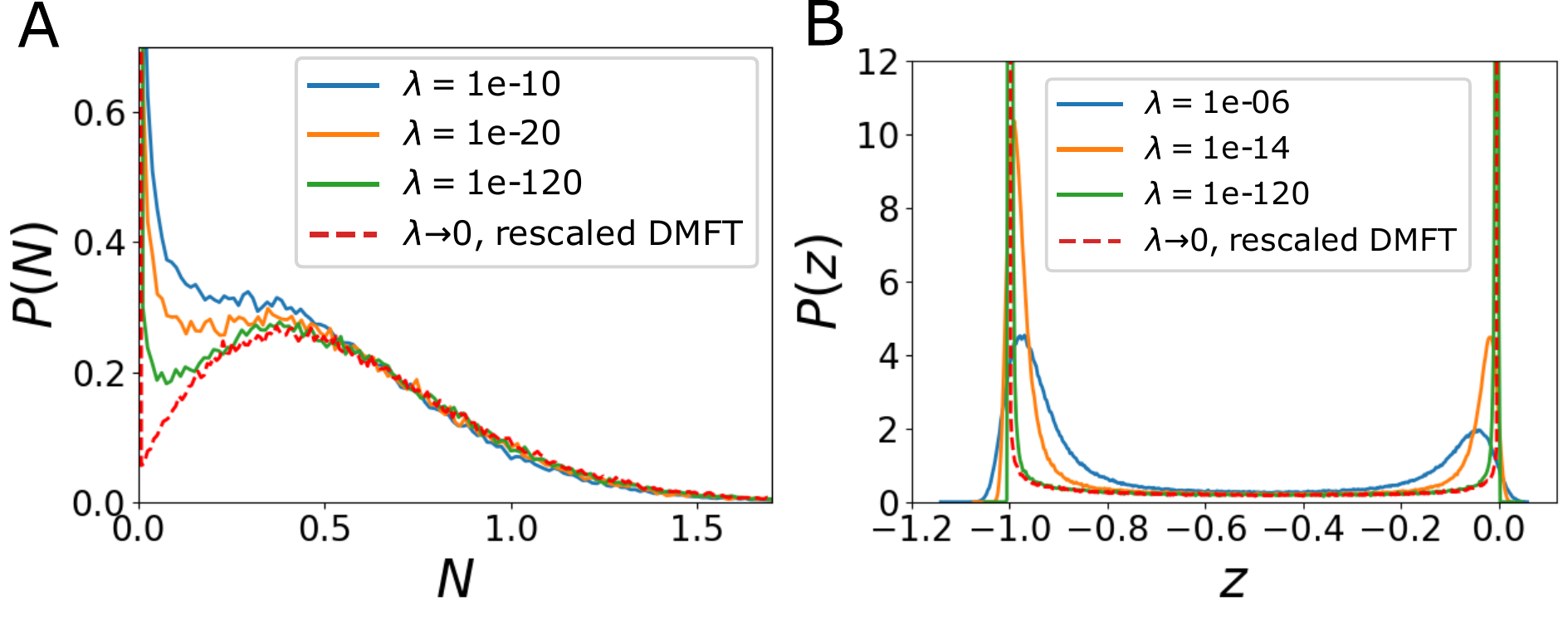}
\par\end{centering}
\caption{\textbf{Collapse of species abundance distributions.} Numerically
measured distributions of $N$ (A) and $z=\ln N/|\ln\lambda|$ (B)
converge to the distributions predicted by the rescaled process as
$\lambda\to0^{+}$.\label{fig:distrib-conv}}
\end{figure}

We now return to the species abundance distribution $P(N)$. As mentioned
in Sec. \ref{subsec:Species-abundance-and}, $P(N)$ behaves roughly
as $1/N$ in the intermediate range $\lambda\ll N\ll1$. Using the
rescaled dynamics, we can refine this statement. The dynamics of $z_{i}(s)$
(Sec. \ref{sec:Rescaled-dynamics}) spend a finite fraction of the
time at the boundaries $z=-1,0$. This translates to two delta-peak
contributions in $P(z)$ at these values. In addition, there is a
regular contribution for $-1<z<0$. Together, this reads
\begin{eqnarray}
P(z) & = & \phi_{\text{top}}\,\delta(z)+\left(1-\phi_{\text{inter}}\right)\delta(z+1)\nonumber \\
 &  & +(\phi_{\text{inter}}-\phi_{\text{top}})h(z)\Theta(-z)\Theta(1+z)\,,\label{eq:distribution_z_mig}
\end{eqnarray}
where $h(z)$ is a smooth function with $\int_{-1}^{0}h(z)\,dz=1$.
$\phi_{\text{top}},\phi_{\text{inter}}$ were defined in Eq. (\ref{eq:phis_from_rescaled}).
Fig. \ref{fig:distrib-conv}(B) shows the collapse of $P(z=\ln N/|\ln\lambda|)$
as $\lambda\to0^{+}$ to this form.

Eq. (\ref{eq:distribution_z_mig}) sets the form of the abundance
distribution in the intermediate range $\lambda\ll N\ll1$. Changing
variables from $z$ to $N$, we get
\begin{equation}
P(N)=\frac{1}{N\left|\ln\lambda\right|}\:h\negthinspace\left(\frac{\ln N}{\left|\ln\lambda\right|}\right)\ .\label{eq:P_N_scaling_corrections}
\end{equation}
This refines the $1/N$ dependence with an additional, slowly-varying
correction. As we discuss in Sec. \ref{sec:Discussion}, this correction
can appear to change the power law exponent of the species abundance
distribution when $\lambda$ is finite, and only parts of the entire
distribution are sampled. The distribution of top species can be inferred
from $P[g|z]$, the distribution of the growth rate $g$ conditioned
on the value of the rescaled abundance $z$. For $N\ge0$, we get

\begin{equation}
P(N)=\phi_{\text{top}}P\left[g=N|z=0\right]+(1-\phi_{\text{top}})\delta(N)\,.\label{eq:distribution_N_mig}
\end{equation}
Fig. \ref{fig:distrib-conv}(A) shows the convergence of the distribution
$P(N)$ to this limiting distribution as $\lambda\to0^{+}$. The results
in the limit $\lambda\to0^{+}$ were obtained by solving numerically
the rescaled DMFT equations, see Eqs. (\ref{eq:DMFT2},\ref{eq:dmft-mean2},\ref{eq:dmft-var2}).
The limiting distribution $P(N)$ deviates from the truncated Gaussian
SAD obtained in the fixed point phase.

\section{Dynamics of an isolated system and transient dynamics at finite $\lambda$\label{sec:Phenomenology:-Epoch-I}}

\subsection{Dynamics of an isolated system}

Isolated systems are characterized by zero migration rate, $\lambda=0$,
which is a singular limit of the Lotka-Volterra system of equations
with a large number of species, in the chaotic phase. Indeed, the
timescale $|\ln\lambda|$ which characterizes these dynamics at finite
$\lambda$, diverges when $\lambda\to0$. For $\lambda=0$, the dynamics
are not time-translation invariant but forever slow down in time,
as evidenced in a linear growth of the correlation time as a function
of the elapsed time, a behavior known in physics as `aging'. Formally,
$C_{\lambda=0}(t,t+tt')\to\hat{C}(t')$ when $t$ is large, see Fig.
\ref{fig:crossover}(A) for the collapse of results from numerical
simulations. Again, the master curve $\hat{C}(t')$, which depends
on the parameters $\sigma$ and $\mu$, is regular at $t'\to0^{+}$.
This behavior can be understood as follows: When $\lambda=0$ the
lowest values of $\ln N$ reached after time $t$ are of order $\ln N\sim-t$.
If a species changes to positive growth rate at this time, it will
therefore take another time $t$ for its population size to be $O(1)$.
This sets the correlation time. A similar mechanism was found in another,
exactly-solvable, model \citep{arnoulxdepirey_Aging_2023}.

The proof of the existence of this aging regime is given in App. \ref{subsec:Rescaled-dynamics-epoch-I}.
Using a reasoning similar to that employed in Sec. \ref{sec:Rescaled-dynamics},
we show that the transformed variables $s\equiv\ln t,z\equiv\ln(N)/t$
obey a well-defined set of DMFT equations which become time-translation
invariant when $t\to\infty$. These transformations reflect a growth
of both timescales and log-fluctuations with the elapsed time. The
resulting process $z(s)$ is different from that of Sec. \ref{sec:Rescaled-dynamics},
and is described in detail in Appendix \ref{subsec:Rescaled-dynamics-epoch-I}.

An important consequence of this result is that the collective deterministic
dynamics in an isolated system drives the population size of any species
arbitrarily close to $0$ as time grows. Considering that actual populations
sizes are finite, integer numbers, this process will inevitably lead
to the extinction of many species, subsequently leading to an arrest
of the fluctuations, as suggested in \citep{roy_Complex_2020,pearce_Stabilization_2020}.

\subsection{Transient dynamics at finite $\lambda$}

When migration is present, this dynamical slowdown provides a mechanism
by which the correlation time grows until time $t_{\textrm{transient}}\sim|\ln\lambda|$,
where the correlation time reaches the value $|\ln\lambda\vert$ discussed
above. This is indeed the time it takes for a finite fraction of the
species in the community to reach the migration floor, when starting
with all species with population sizes of $O(1)$, and before which
$\lambda$ can be safely set to zero.

\begin{figure}[!tb]
\begin{centering}
\includegraphics[width=1\columnwidth]{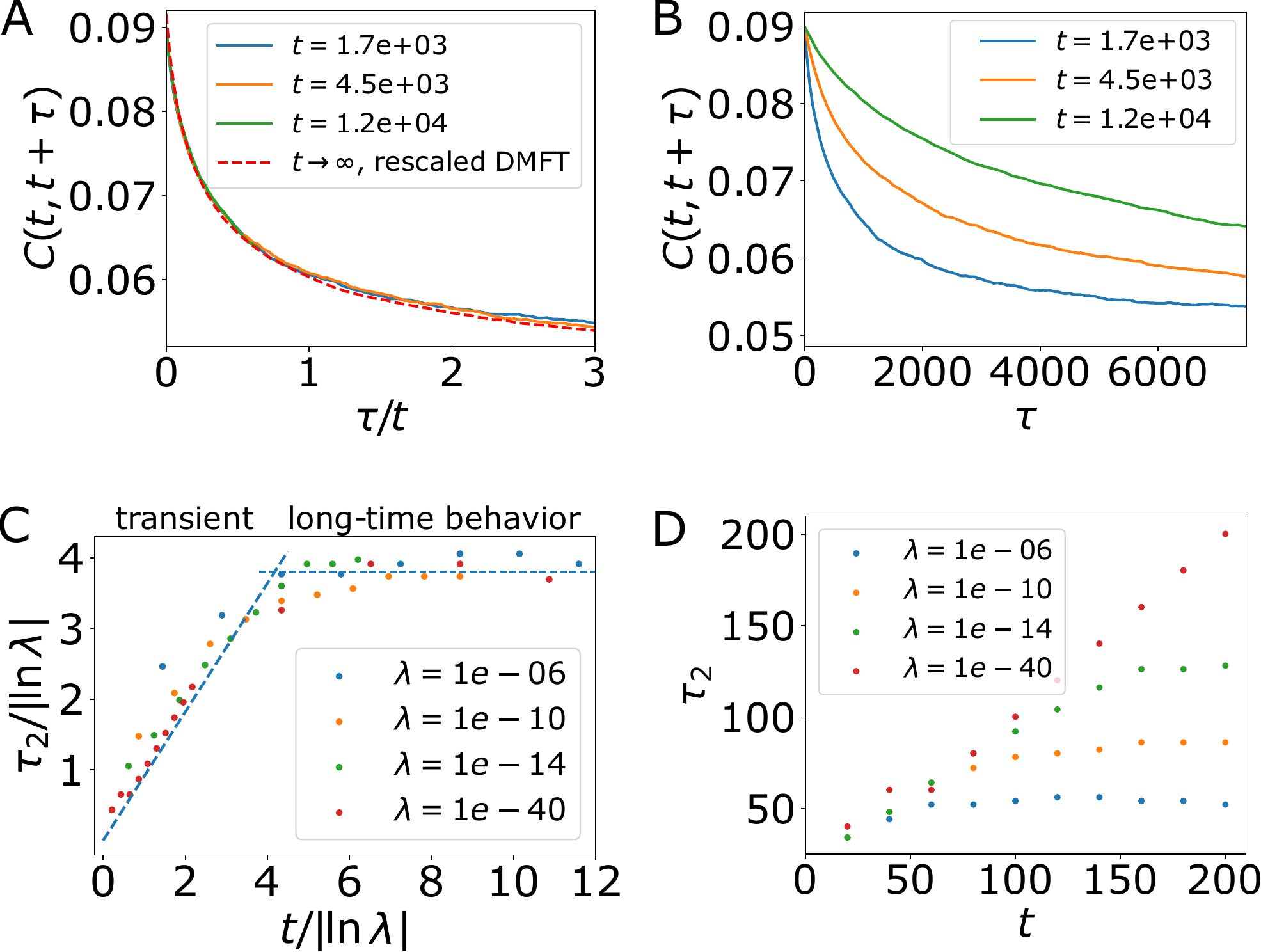}
\par\end{centering}
\caption{\label{fig:crossover}\textbf{Growth of timescales in dynamics without
migration.} (A,B) When $\lambda=0$, the collapse of $C(t,t+\tau)$
as a function of $\tau/t$ demonstrates the linear growth of timescales
with the elapsed time. Data without rescaling, as a function only
of $\tau$, is shown for comparison in (B). (C,D)\textbf{ }Crossover
at finite $\lambda$ from a transient regime exhibiting the $\lambda=0$
phenomenology to the long-time behavior, as identified by the time
constant $\tau_{2}$ of the correlation function.\textbf{ }After an
initial transient of finite time duration, \textbf{$\tau_{2}$} grows
linearly with the elapsed time and independently of $\lambda$, a
trademark of the $\lambda=0$ dynamics. The crossover to the long-time
behavior happens around times proportional to $|\ln\lambda|$, after
which $\tau_{2}$ stabilizes to a value proportional to $\left|\ln\lambda\right|$.
This scaling of the crossover is manifest in the data collapse in
(C), with the data without rescaling displayed in (D) for comparison.
$\tau_{2}(t,\lambda)$ is defined as the time when $\left[C(t,t+\tau_{2})-C(t,\infty)\right]/\left[C(t,t)-C(t,\infty)\right]=e^{-1}$.
The value of $C(t,\infty)$ is estimated through an exponential fit
of the function $C(t,t+\tau)$ as a function of $\tau$.}
\end{figure}

We verify in simulations that the transient dynamics is characterized
by a linear growth of timescale with the elapsed time, which is interrupted
at a time $t_{\textrm{transient}}\sim|\ln\lambda|$. For that, we
measure the time constant $\tau_{2}(\lambda,t)$ it takes for the
autocorrelation function $C_{\lambda}(t+\tau,t)-C_{\lambda}(\infty,t)$
to reach a fraction $e^{-1}$ of its $\tau=0$ value, starting at
initial time $t$ and with migration rate $\lambda$. Fig. \ref{fig:crossover}(C)
shows that the growth and saturation of the timescale follow the scaling
relation
\[
\frac{\tau_{2}(\lambda,t)}{|\ln\lambda|}=f\left(\frac{t}{|\ln\lambda|}\right)\,,
\]
where $f$ is a smooth function with $f(x)\propto x$ at small $x$,
encoding the slowdown of the dynamics in the transient regime, and
$f(x)$ approaching a constant as $x\to\infty$, encoding the time-translation
invariant behavior of the long-time dynamics, with correlation timescale
$|\ln\lambda|$.

\section{Robustness of the predictions\label{sec:robustness}}

The theory is built around two limits, $S\to\infty$ and $\lambda\to0^{+}$,
and assumes that the interaction coefficients $\alpha_{ij}$ are sampled
independently. In this last part of this work, we assess the robustness
of our predictions when these assumptions are relaxed. We find that
the main qualitative features discussed in this work are robust against
changes in model definition, and relevant even at reasonable values
of migration rate and number of species.

\subsection{Finite migration rate $\lambda$ and number of species $S$}

We start by discussing how the key quantities, $\left\langle N\right\rangle $,
$\left\langle N^{2}\right\rangle $ and $\phi_{{\rm top}}$ vary with
$S,\lambda$, when measured from abundance data gathered in numerical
simulations of the original many-species dynamics Eq. (\ref{eq:multibody}).
Regarding the moments $\left\langle N\right\rangle ,\left\langle N^{2}\right\rangle $,
we find that the dependence on $S$ and $\lambda$ is very weak for
$\left\langle N\right\rangle $ and weak for $\left\langle N^{2}\right\rangle $,
respectively within roughly 1\% and 10\% in the inspected range of
parameter, see Fig. \ref{fig:N_N2} in Appendix \ref{sec:mean_var}.

\begin{figure}[!tb]
\begin{centering}
\includegraphics[width=1\columnwidth]{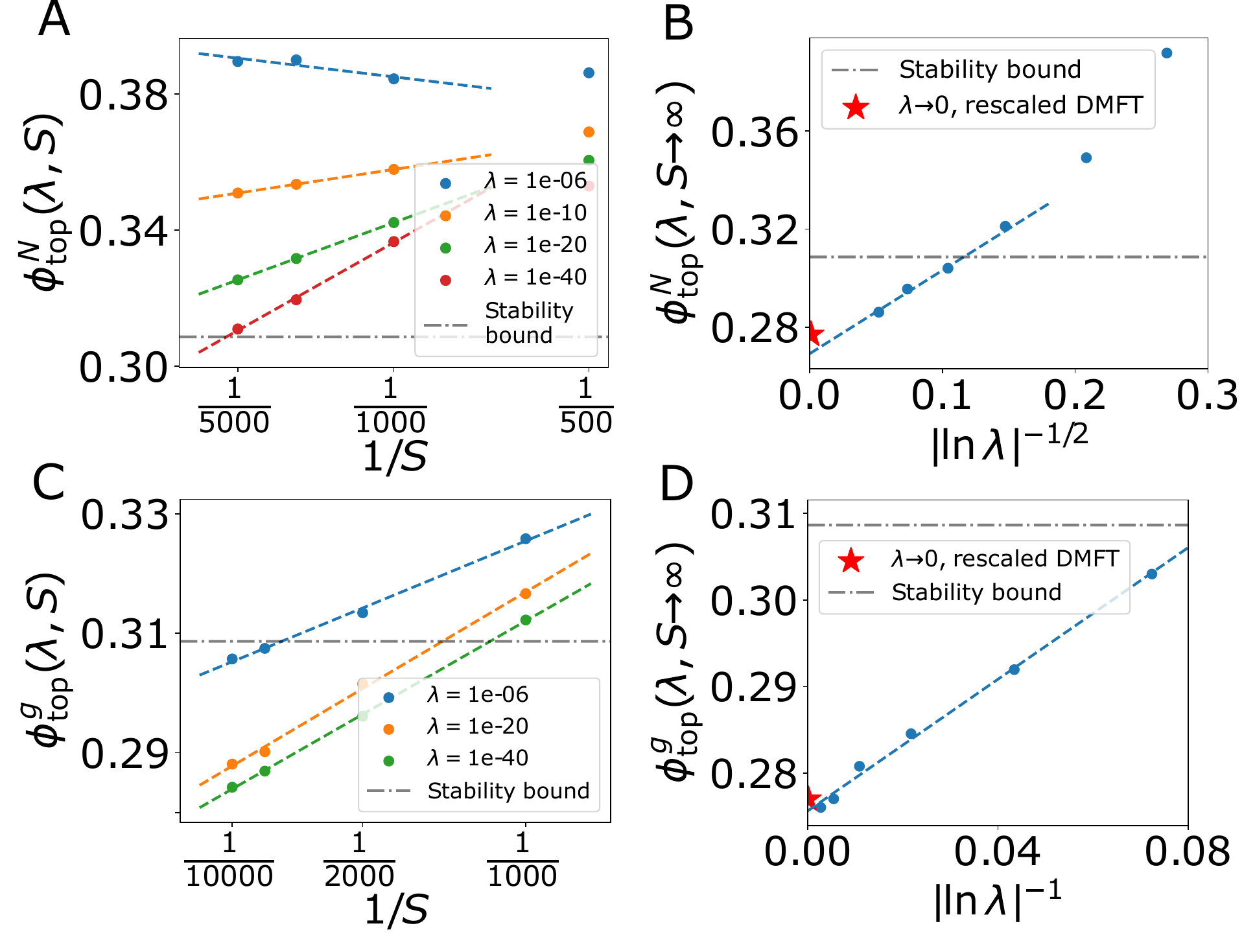}
\par\end{centering}
\caption{\label{fig:abundance_filtering-1-1}\textbf{Average top diversity
measures, $\phi_{{\rm top}}^{\text{N}}(S,\lambda,\epsilon),\phi_{{\rm top}}^{\text{g}}(S,\lambda,\epsilon)$,
as a function of $S$ and $\lambda$.} $\phi_{{\rm top}}^{\text{N}}$
is the fraction of species with $N_{i}>\epsilon=10^{-3}$; $\phi_{{\rm top}}^{\text{g}}$
also requires $g_{i}>0$. (A) $\phi_{{\rm top}}^{N}(S,\lambda,\epsilon)$
as a function of $S$ converges as $\phi_{{\rm top}}^{N}(S\to\infty,\lambda,\epsilon)+\#/S$,
at fixed $\lambda$. Data points at $S=500$ are excluded from the
linear fit. (B) Large $S$ value of the top diversity, $\phi_{{\rm top}}^{N}(S\to\infty,\lambda,\epsilon)$,
converges as $\left|\ln\lambda\right|^{-1/2}$ to its asymptotic value
$\phi_{{\rm top}}^{N}(S\to\infty,\lambda\to0^{+},\epsilon)$. The
asymptotic value $\phi_{{\rm top}}^{N}(S\to\infty,\lambda\to0^{+},\epsilon)$
is within 5\% of the value predicted by the rescaled DMFT, and well
below the stability bound. For all reasonable values of the migration
rate, (here already when $\lambda\gtrsim10^{-40}$), the measured
top diversity is significantly above the stability bound. $\phi_{{\rm top}}^{N}(S\to\infty,\lambda,\epsilon)$
was inferred from the finite $S$ scaling, panel A. (C,D) Same as
panels A,B, but for $\phi_{{\rm top}}^{\text{g}}(S,\lambda,\epsilon)$.
The convergence to the asymptotic $\phi_{{\rm top}}$ is more rapid;
in particular, $\phi_{{\rm top}}^{\text{g}}$ can be either below
or above the stability bound.}
\end{figure}

To measure $\phi_{{\rm top}}$ at finite $\lambda$, we test three
options. The first measure, $\phi_{{\rm top}}^{\text{N}}(S,\lambda,\epsilon)$,
is simply a threshold on the abundance: counting all the species with
$N_{i}\geq\epsilon$ at a given time, for some chosen $\epsilon$.
Unsurprisingly, $\phi_{{\rm top}}^{\text{N}}(S,\lambda,\epsilon)$
is more sensitive to $\lambda$ than $\left\langle N\right\rangle $
or $\left\langle N^{2}\right\rangle $. For moderate values of $S$
and $\lambda$, we find that the asymptotic value $\phi_{{\rm top}}$
nevertheless provides a reasonable estimate of $\phi_{{\rm top}}^{\text{N}}(S,\lambda,\epsilon)$,
see Fig. \ref{fig:abundance_filtering-1-1}(A, B), though the discrepancy
increases with the strength of the interactions, see Fig. \ref{fig:large_params}(B).
This last point suggests that disentangling species at high and intermediate
abundances becomes harder when the scale of the interactions, and
thus the amplitude of the fluctuations, increases. The second measure,
denoted by $\phi_{{\rm top}}^{\text{g}}(S,\lambda,\epsilon)$, refines
the first one and requires $N_{i}\geq\epsilon$ but also $g_{i}>0$.
We find that both measures converge to the asymptotic $\phi_{{\rm top}}$
as $1/S$ in $S$. Yet the convergence with $\lambda$ is significantly
different, with $\phi_{{\rm top}}^{\text{N}}$ converging as $\left|\ln\lambda\right|^{-1/2}$,
and $\phi_{{\rm top}}^{\text{g}}$ faster, as $\left|\ln\lambda\right|^{-1}$,
see Fig. \ref{fig:abundance_filtering-1-1}(B, D). This highlights
the relevance of the growth rate $g$ in determining ``top species'',
namely members of fixed points. We find that the measure $\phi_{{\rm top}}^{\text{N}}(S,\lambda,\epsilon)$
is often above the stability bound, in contrast to the asymptotic
$\phi_{{\rm top}}$ which is always below it. Due to the faster convergence,
$\phi_{{\rm top}}^{\text{g}}(S,\lambda,\epsilon)$ can be either above
or below this bound, depending on the parameters, see Fig. \ref{fig:abundance_filtering-1-1}(C).
The origin of these convergence rates is discussed in Appendix \ref{sec:Additional-to-discussion}.

Finally, we consider a situation where one is able to manipulate the
system, by removing certain species and continuing the dynamics. Namely,
after a long time so that transients have passed, we kill all species
with $N_{i}/g_{i}<1/2$. These correspond to species with positive
growth rate that are in the midst of their jump, below halfway, as
well as those with negative growth rate. (Recall that, asymptotically
as $\lambda\to0^{+}$, abundant species are those for which that $N_{i}/g_{i}=1$.)
Then, we run again the dynamics, and we find that the remaining species
reach a stable equilibrium. The properties of the equilibrium obtained
in this way are strikingly similar to those predicted by the asymptotic
theory, even for reasonable $\lambda$. The asymptotic diversity and
the asymptotic distribution in Eq. (\ref{eq:distribution_N_mig})
can be almost exactly reproduced, see Appendix \ref{sec:id_top}.

\subsection{Correlation between the matrix elements}

In addition to taking the asymptotic limits in $S,\lambda$, the model
above assumes that interactions are statistically asymmetric with
vanishing correlation coefficient $\text{corr}(\alpha_{ij},\alpha_{ji})=0$.
We show that two of our main qualitative results--how the timescale
grows and that top diversity is below the stability bound--also hold
when this assumption is relaxed. Allowing for more symmetric or anti-symmetric
interactions, we take a correlation coefficient $\text{corr}(\alpha_{ij},\alpha_{ji})=\gamma$
with $-1<\gamma<1$.

The growth of the timescale $\tau\sim\tau_{\lambda}\equiv\left|\ln\lambda\right|$
is clearly seen in Fig. \ref{fig:correlation_gamma}(A,B) in Appendix
\ref{subsec:Generalize-gamma}, the equivalent of Fig. \ref{fig:dynamics-1}(C).
This is expected for the same reason as when $\gamma=0$: the time
for a population to grow from $\lambda$ to $N\sim1$ scales as $\left|\ln\lambda\right|$.
As to the diversity $\phi_{\text{top}}$, we conjecture that the slowness
of the dynamics at $\lambda\ll1$ still introduces a clear partition
between nearly-extinct species with $\ln N\sim\ln\lambda$ and abundant
species with $N=O(1)$. For the abundant species, the long timescale
implies that they are near a fixed point. In Appendix \ref{subsec:Generalize-gamma},
we generalize the relation between fluctuations and diversity from
Sec. \ref{subsec:Diversities-revisited}, Eq. (\ref{eq:fluct_div}),
to give

\begin{equation}
G=2|\hat{C}'\left(0^{+}\right)|\ \frac{\left(1+\sqrt{1-4\gamma\phi_{{\rm top}}\sigma^{2}}\right)^{2}-4\phi_{{\rm top}}\sigma^{2}}{\left(1+\sqrt{1-4\gamma\phi_{{\rm top}}\sigma^{2}}\right)^{2}}\,.\label{eq:rel_gamma_main-1}
\end{equation}
As for $\gamma=0$, we expect that the growth of timescale when $\lambda\to0^{+}$
goes hand-in-hand with jumps in the rescaled dynamics of the abundances,
meaning $G>0$. This implies $\left(1+\sqrt{1-4\gamma\phi_{{\rm top}}\sigma^{2}}\right)^{2}-4\phi_{{\rm top}}\sigma^{2}>0$,
so that $\phi_{{\rm top}}$ lies strictly below the linear stability
bound \citep{bunin_Ecological_2017}. This is confirmed by numerical
simulations, see Fig. \ref{fig:correlation_gamma}(C,D) in Appendix
\ref{subsec:Generalize-gamma}.

\section{Discussion\label{sec:Discussion}}

We begin the discussion by summarizing key predictions presented above.
We identified several signatures of the many-species Lotka-Volterra
dynamics, when the abundnaces does not reach a fixed point and migration
rates are small. They may serve as footprints of endogenously-driven
fluctuations in experimental or natural situations.

First, we predict that endogenously-driven fluctuations in time would
lead to broadly distributed abundance distributions, in contrast with
the fixed point distributions. For very low migration, the abundance
distribution at intermediate values (meaning small population sizes
well above the migration floor) are predicted to behave as $P(N)\sim1/N$.
At any finite $\lambda$, there are slowly varying corrections to
this $1/N$ behavior, see Eq. (\ref{eq:P_N_scaling_corrections}).
As $P(N)$ is broad, one can always define a slowly-varying `local'
power law $\nu(z)$ by the slope of $\ln P(N)$ versus $\ln N$ at
given $z=\ln N/\left|\ln\lambda\right|$. It has corrections of order
$|\ln\lambda|^{-1}$, $\nu(z)=-1+|\ln\lambda|^{-1}h'(z)/h(z)$, which
might explain deviations from $\nu=-1$ in observed abundance distributions
\citep{ser2018ubiquitous}. These corrections vary with $z$ and are
arbitrarily large over the entire range $-1<z<0$, so no unique exponent
other than $-1$ can be defined over the entire range of $N$.

Second, we predict that a single timescale controls the dynamics,
predicted to grow as $|\ln\lambda|$ when lowering the migration rate.
This could be tested in controlled experiments, for example via the
abundance autocorrelation.

Moving on to diversity, different definitions of diversity give different
results. For example, the number of species that can grow from rare
at a given time, is generally different from the number of species
that have high abundance, in contrast with the situation at a fixed
point. We show that the number of species with high, intermediate
and low abundance is well defined, namely insensitive to the precise
threshold above (or below) which the number of species is counted.
We show that the number of species at high abundance is significantly
below the May bound (their fraction of the total number of species
is below the fraction allowed by the bound). This makes the community
of high-abundance species a stable equilibrium of the dynamics if
the other species are removed. The distance to the stability bound
increases with the strength of the interactions $\sigma$, see Fig.
\ref{fig:richness-SAD-1}(G), and also Fig. \ref{fig:large_params}(B)
in Appendix \ref{sec:Additional-to-discussion}. In contrast, when
including the intermediate ones (in the power law regime of $P(N)$),
the total number goes above the bound. The same is true when including
all species in the pool that may invade. It is this last fact that
drives species turnover: there are always species that grow from rare
to replace the ones at high abundance.

\begin{figure}[!tb]
\begin{centering}
\includegraphics[width=1\columnwidth]{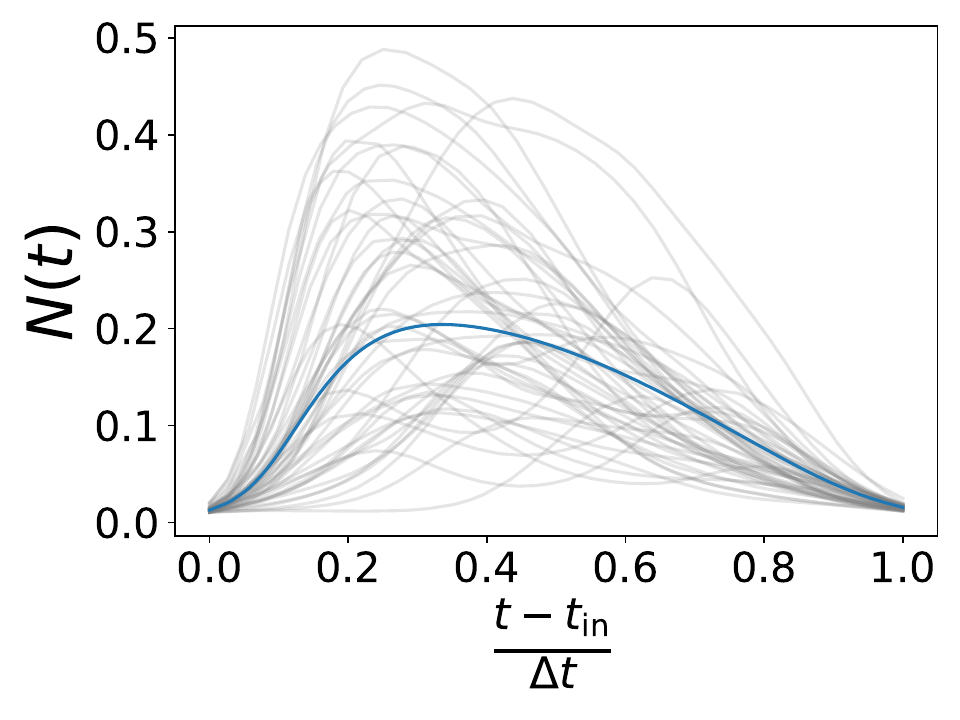}
\par\end{centering}
\caption{\label{fig:finite-migration-Trajectories}\textbf{Asymmetry of blooms
under time-reversal.} Trajectories of $N(t)$ while at high-abundance
display a clear asymmetry in time, with a rapid initial increase and
more gradual decrease. This is conspicuous even for migration rates
that are not very small (here $\lambda=10^{-6}$). It is the finite-$\lambda$
counterpart of the asymptotic behavior at $\lambda\to0^{+}$, featuring
sharp jumps from zero to positive $N$, see Fig. \ref{fig:The-rescaled-dynamics.}.
Shown are trajectories that go above $N=10^{-2}$ at some time $t_{\text{in}}$
and stay above it until time $t_{\text{in}}+\Delta t$, and reach
$N\ge0.1$ at some intermediate time. Here $\Delta t\simeq5.8\left|\ln\lambda\right|$
that is the most common length of such trajectories. In light grey
are 22 example trajectories, and the thick blue line shows the average
over many such trajectories.}
\end{figure}

Last, a key property of the dynamics at low migration is the existence
of jumps from rare in the dynamics of population sizes, see Fig. \ref{fig:The-rescaled-dynamics.}.
For finite $\lambda$, this manifests itself in a strong asymmetry
of ``blooms'', namely trajectories where the abundance of a species
increases from rare before returning there. This can be clearly observed
in time series at finite $\lambda$, see Fig. \ref{fig:finite-migration-Trajectories},
and would be very interesting to look for in experimentally-measured
time series.

In conclusion, the dynamically-fluctuating phase of high-diversity
ecological communities is a promising direction to explain key features
of natural high-diversity ecosystems. We offered a list of additional
predictions expected in this phase. It would be interesting to further
investigate the robustness of these features upon modifying the structure
of the interaction matrix. Along this line, we note that the chaotic
dynamics appear at finite $S$ also without any beneficial interactions
($\alpha_{ij}<0$) between species, and the analysis above is expected
to hold. We further note that the limit of large $\sigma,\mu$ in
our framework connects to other asymptotic limits \citep{pearce_Stabilization_2020,lorenzana2022well}.
A recent work on the strongly interacting case \citep{mallmin2023chaotic}
suggests from numerics that the qualitative picture of the present
study may extend to that regime, in particular the existence of a
growing timescale, and the fact that dynamics evolve in the vicinity
of fixed points. A different and very interesting direction for future
research is understanding how these results extent to spatially-extended
metacommunities \citep{roy_Complex_2020,pearce_Stabilization_2020},
beyond a constant migration from an unspecified species pool, as assumed
here. This question is pertinent, given that the present work shows
that chaotic fluctuations cannot generically be sustained in isolated
high-diversity systems, due to extinctions.

\medskip{}

\emph{Acknowledgments}--This work was supported by the Israel Science
Foundation (ISF) Grant No. 773/18.\clearpage{}

\appendix\onecolumngrid

\section{Derivations}
\subsection{Regularity of the correlation function as $\lambda\to0^{+}$
\label{subsec:regularity}}

\begin{figure}[H]
\begin{centering}
\includegraphics[width=0.25\columnwidth]{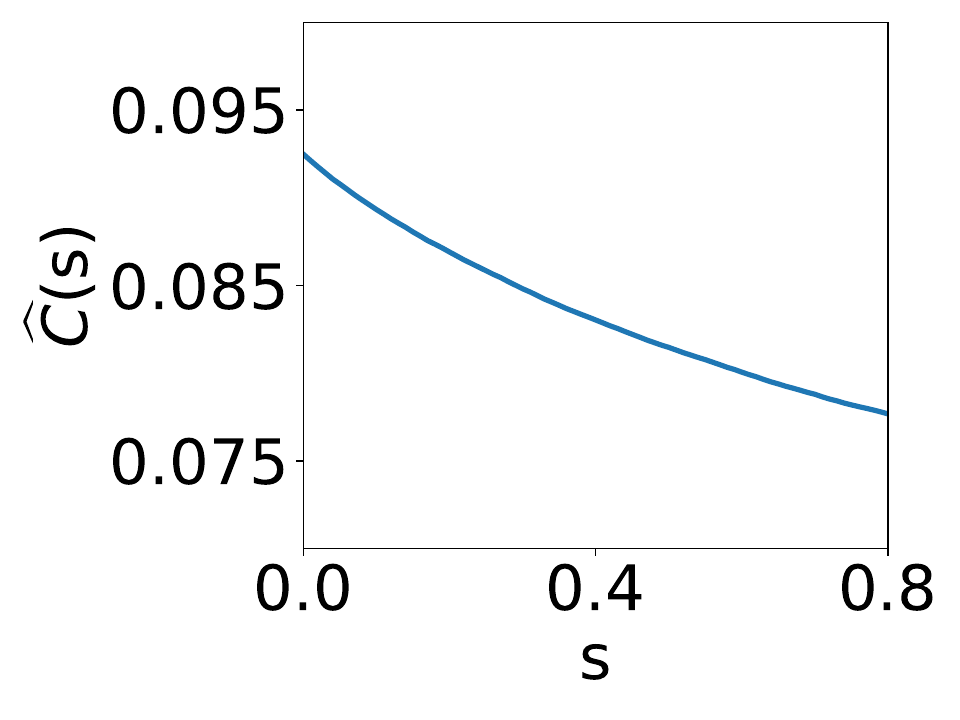}
\par\end{centering}
\caption{\label{fig:der_c}\textbf{Zoom on the small $s$ behavior of $\hat{C}(s)$.
}The correlation function was obtained by numerically solving the
DMFT equations Eqs. (\ref{eq:DMFT2},\ref{eq:dmft-mean2},\ref{eq:dmft-var2}).
The derivative is finite at $s=0^{+}$, thereby confirming the absence
of fast-time scale in the Lotka-Volterra dynamics.}
\end{figure}

By solving the DMFT equations Eqs. (\ref{eq:DMFT2},\ref{eq:dmft-mean2},\ref{eq:dmft-var2})
following the procedure detailed in App. (\ref{sec:Numerical-methods}),
we find that $\hat{C}'(0^{+})$ is finite, thus confirming the absence
of fast-time scale in the Lotka-Volterra dynamics, see Fig. \ref{fig:der_c}.

\subsection{Rescaled dynamics for $\lambda=0$\label{subsec:Rescaled-dynamics-epoch-I}}

In this section, we adapt the derivation of Sec. \ref{sec:Rescaled-dynamics}
to the singular case $\lambda=0$ that accounts for the initial transient
when $0<\lambda\ll1$. For $0<\lambda\ll1$, the system reaches a
time-translation invariant state with correlations characterized by
a unique time-scale $\vert\ln\lambda\vert$. The latter diverges as
$\lambda\to0^{+}$, suggesting that the $\lambda=0$ dynamics does
not reach a time-translation invariant state. We show that the dynamics
age with a correlation time that grows linearly with the elapsed time,
a phenomena already identified in a related population dynamics model
\citep{arnoulxdepirey_Aging_2023},
\begin{equation}
\lim_{t\to\infty}C_{\lambda=0}(t,t\text{e}^{s})=\hat{C}(s)\,.\label{eq:scaling_aging_DMFT}
\end{equation}
This scaling regime can be shown to be a self-consistent solution
of the dynamical mean-field theory equations Eqs. (\ref{eq:DMFT1},\ref{eq:dmft_mean1},\ref{eq:dmft-var1})
with $\lambda=0$. We introduce $z\equiv\ln N/t$ and $s\equiv\ln t$
and obtain from Eq. (\ref{eq:DMFT1})

\[
z'(s)=-z(s)+g(s)-\exp(e^{s}z(s))\,.
\]
Under the assumption that Eq. (\ref{eq:scaling_aging_DMFT}) holds,
the Gaussian process $g(s)$ has a time-independent mean and finite
memory with time-translation invariant correlations in the large $s$
limit. Similarly to the $\lambda\to0^{+}$ case, the term $-\exp(e^{s}z)$
effectively acts as a hard wall at $z=0$ thus constraining $z(s)\leq0$.
The long-time dynamics therefore writes
\begin{equation}
z'(s)=-z(s)+g(s)-W(z),\label{eq:DMFT_aging}
\end{equation}
with $W(z)$ formally accounting for the presence of the confining
boundary. We use the non-penetrability condition to resolve the ambiguous
expression $N(s)=\exp(e^{s}z)$ in the double limit $s\to\infty$
and $z\to0$ and get 
\begin{equation}
N(s)\equiv W(z(s))=g(s)\Theta(z(s))\,,\label{eq:non_penetrability_aging}
\end{equation}
with the convention $\Theta(0)=1$. It shows that Eq. (\ref{eq:DMFT_aging})
is supplemented by the same self-consistency conditions as in the
$\lambda\to0^{+}$ case, see Eqs. (\ref{eq:dmft-mean2},\ref{eq:dmft-var2})

\begin{equation}
\langle g(s)\rangle=1-\mu\langle g(s)\Theta(z(s))\rangle\,,\label{eq:dmft_mean_aging}
\end{equation}
together with

\begin{equation}
\langle g(s)g(s')\rangle-\langle g(s')\rangle\langle g(s)\rangle=\sigma^{2}\langle g(s)g(s')\Theta(z(s))\Theta(z(s'))\rangle\,.\label{eq:selfconsz_aging}
\end{equation}
Note however that in the $\lambda=0$ case, the process $z(s)$ is
confined in the negative half-line by a harmonic potential and not
by a hard boundary, see Eq. (\ref{eq:DMFT_aging}).

\begin{figure}[!tb]
\begin{centering}
\includegraphics[width=0.5\columnwidth]{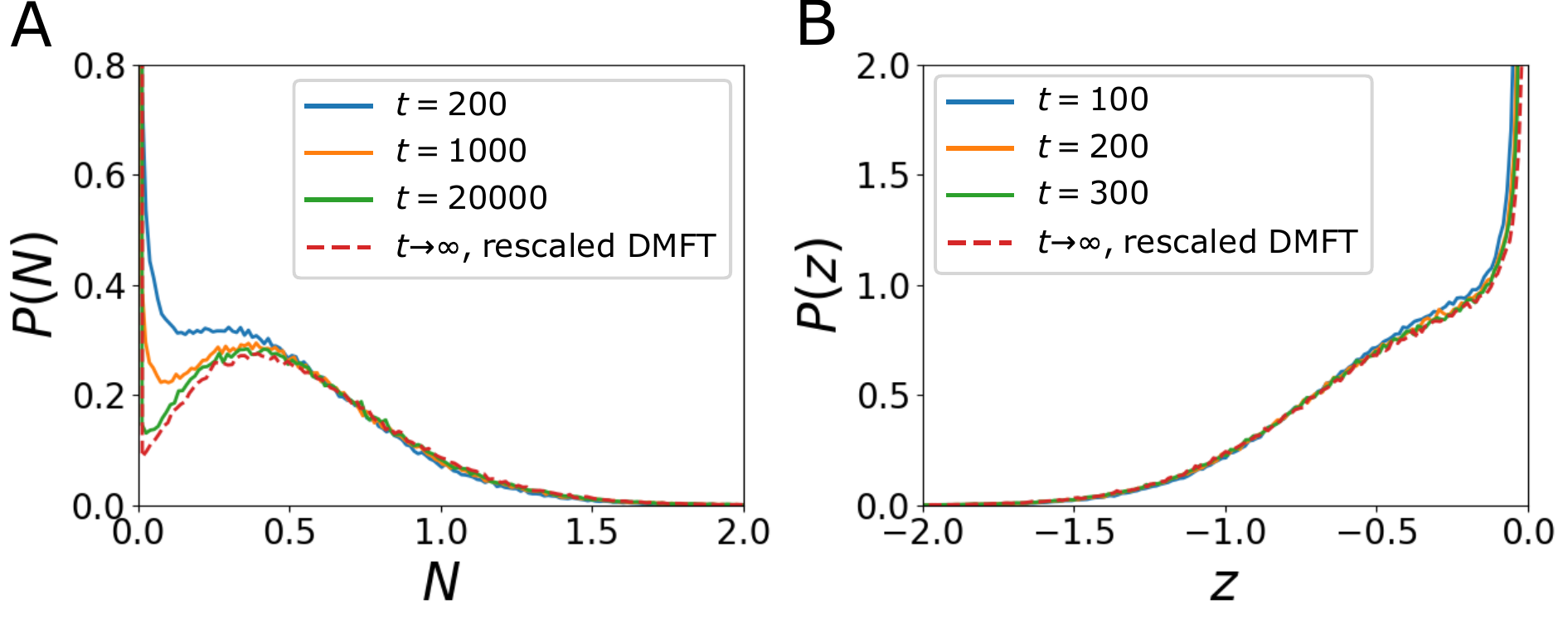}
\par\end{centering}
\caption{\textbf{Collapse of species abundance distributions.} Numerically
measured distributions of $N$ (A) and $z=\ln N/t$ (B) converge to
the distributions predicted by the rescaled process as $t\to\infty$.\label{fig:distrib-conv-1}}
\end{figure}

Under the condition that $g(s)$ has time-translation invariant correlations,
Eq. (\ref{eq:DMFT_aging}) manifestly predicts that $z(s)$ reach
at long time a time-translation invariant state. The closure equations
Eqs. (\ref{eq:dmft_mean_aging})-(\ref{eq:selfconsz_aging}) then
self-consistently show the validity of the time-translation invariant
ansatz in rescaled time $s$, eventually showing the validity of the
aging scaling given in Eq. (\ref{eq:scaling_aging_DMFT}). In Fig.
\ref{fig:distrib-conv-1}, we show the large-time convergence of the
distributions $P(N)$ and $P(z)$ to those predicted by the $\lambda=0$
rescaled dynamics presented here.

\subsection{Diversity limited by fluctuations\label{subsec:Diversity-limited-by}}

The two cases $\lambda\to0^{+}$ and $\lambda=0$ share a crucial
property: the long-time dynamics are very slow, reflecting the fact
that the system evolves in the vicinity of feasible fixed points (all
population sizes are $\geq0$). More precisely, at any given time
$s$, the system is close to a fixed point comprised of some abundant
species with $O(1)$ population sizes (corresponding to the fraction
$\phi_{{\rm top}}$ of species with $z(s)=0$), and some rare species
(corresponding to the fraction $1-\phi_{{\rm top}}$ of species with
$z(s)<0$). Species turnover happens because these fixed points are
invadable, meaning that some nearly-extinct species have positive
growth rate. Furthermore, we see that the subset of abundant species,
when taken alone, is linearly stable and not marginal. Indeed, the
fraction $\phi_{{\rm top}}$ of top species does not saturate the
stability bound with $\phi_{{\rm top}}\sigma^{2}<1$, see Fig. \ref{fig:richness-SAD-1}(G).
Despite the fact that abundant species are found in the vicinity of
a stable fixed point, the dynamics exhibit abundance fluctuations
due to the continuous flux of incoming species from the pool of nearly
extinct ones.

Here we derive a relation between temporal fluctuations and the observed
diversity valid both in the aging ($\lambda=0$) and chaotic (for
$\lambda\to0^{+}$) regimes of the many-body Lotka-Volterra system
of equations that then establishes the linear stability of the subset
of abundant species $\phi_{{\rm top}}\sigma^{2}<1$, see Eq. (\ref{eq:fluct_div}).
We take advantage of the slowness of the dynamics to generalize the
calculation of the fluctuations induced by a random perturbation to
a fixed point \citep{bunin_Ecological_2017}. Between the times $s$
and $s+{\rm d}s$, new species become abundant and induce a perturbation
on the species already abundant at time $s$. We then relate the fluctuations
of their population sizes to the amplitude of the effective perturbing
field and conclude by using the DMFT closure relations. Henceforth
we use the notations $\Theta=\Theta(z(s))$ and for any quantity $x(s)$
we write $x(s)=x$ and $\delta x=x(s+ds)-x(s)$. Using $N(s)=g(s)\Theta(z(s))$,
we obtain
\[
\delta N=(\Theta+\delta\Theta)\delta g+g\delta\Theta\,.
\]
Hence,

\begin{align}
\lim_{\mathrm{d}s\to0}\left\langle \frac{\left(\delta N\right)^{2}}{\mathrm{d}s}\right\rangle = & \lim_{\mathrm{d}s\to0}\left\langle \left(\Theta+2\Theta\delta\Theta+\left(\delta\Theta\right)^{2}\right)\frac{\left(\delta g\right)^{2}}{\mathrm{d}s}\right\rangle +\lim_{\mathrm{d}s\to0}\left\langle g^{2}\frac{\left(\delta\Theta\right)^{2}}{\mathrm{d}s}\right\rangle +\lim_{\mathrm{d}s\to0}2\left\langle \left(\Theta+\delta\Theta\right)\delta\Theta g\frac{\delta g}{\mathrm{d}s}\right\rangle \,.\nonumber \\
\label{eq:growth_fluct_app}
\end{align}

We assume that over short times intervals $\mathrm{d}s$, the changes
in $g(s)$ scale to leading order as $\delta g\sim\sqrt{\mathrm{d}s}$,
as for Brownian motion. Furthermore, $\delta\Theta\in\{-1,0,1\}$
and its moments scale as ${\rm O(d}s)$. To prove the latter, we evaluate
the mean number of species with $\delta\Theta=-1$, namely the mean
number of species going from $z<0$ to $z=0$ between $s$ and $s+{\rm d}s$
(which is also the mean number of species going from $z=0$ to $z<0$
in the same time interval). Denoting $\mathbb{P}[g,z]$ the steady-state
joint probability of $g$ and $z$, it reads
\[
\left\langle \left(\Theta-1\right)\delta\Theta\right\rangle =\int_{-\infty}^{0^{-}}\mathrm{d}z\int_{0}^{+\infty}\mathrm{d}g\,\mathbb{P}[g,z]\Theta(z+g\mathrm{d}s)=\mathrm{d}s\int_{0}^{+\infty}\mathrm{d}g\,\mathbb{P}[g,z=0^{-}]g\,,
\]
which indeed scales as $O({\rm d}s)$. We can now treat separately
all the terms appearing in Eq. (\ref{eq:growth_fluct_app}). First,
\[
\lim_{\mathrm{d}s\to0}\left\langle \left[2\Theta\delta\Theta+\left(\delta\Theta\right)^{2}\right]\frac{\left(\delta g\right)^{2}}{\mathrm{d}s}\right\rangle =0\,,
\]
since the term in brackets scales as ${\rm O(d}s)$. Second, if $\Theta\delta\Theta\neq0$,
meaning $z(s)=0$ and $z(s+{\rm d}s)<0$, then we must have $g\sim\sqrt{\mathrm{d}s}$
so that $g(s+{\rm d}s)<0$ might be obtained. This guarantees that

\[
\lim_{\mathrm{d}s\to0}\left\langle \Theta\delta\Theta\,g\frac{\delta g}{\mathrm{d}s}\right\rangle =0\,.
\]
Additionally, by using the identity $\left(\delta\Theta\right)^{2}=-\Theta\delta\Theta+(1-\Theta)\delta\Theta$,
we get
\begin{align*}
\lim_{\mathrm{d}s\to0}\left\langle \left(\delta\Theta\right)^{2}g\frac{\delta g}{\mathrm{d}s}\right\rangle  & =\lim_{\mathrm{d}s\to0}\left\langle -\Theta\delta\Theta\,g\frac{\delta g}{\mathrm{d}s}\right\rangle +\lim_{\mathrm{d}s\to0}\left\langle (1-\Theta)\delta\Theta\,g\frac{\delta g}{\mathrm{d}s}\right\rangle \,,\\
 & =\lim_{\mathrm{d}s\to0}\left\langle (1-\Theta)\delta\Theta\,g\frac{\delta g}{\mathrm{d}s}\right\rangle \,,
\end{align*}
where we used the above result to obtain the last equality. We note
that a generic time-translation invariant Gaussian process can be
generated from the Langevin equation,

\[
\xi'(s)=-k\xi+\sqrt{2D}\eta(s)+\sqrt{2D}\int_{0}^{+\infty}\mathrm{d}s'\,J(s')\eta(s-s')\,,
\]
with $\eta(s)$ a Gaussian white noise and $J(s)$ a suitably chosen
memory kernel that enforces $\left\langle \xi(s)\xi(s')\right\rangle =C(s-s')$,
\emph{i.e. }in Fourier space\emph{ }(with $J(s<0)=0$)\emph{
\[
\hat{C}(\omega)=\frac{2D}{k^{2}+\omega^{2}}|1+\hat{J}(\omega)|^{2}\,.
\]
}This means that the $O(\sqrt{ds})$ increments of $\delta g$ are
statistically independent from the previous history of the system.
We can therefore write 
\[
\lim_{\mathrm{d}s\to0}\left\langle (1-\Theta)\delta\Theta\,g\frac{\delta g}{\mathrm{d}s}\right\rangle =0\,,
\]
because in the average $\delta g$ scales as $O({\rm d}s)$. Lastly,
we have
\[
\lim_{\mathrm{d}s\to0}\left\langle \Theta\frac{\left(\delta g\right)^{2}}{\mathrm{d}s}\right\rangle =\phi_{{\rm top}}\lim_{\mathrm{d}s\to0}\left\langle \frac{\left(\delta g\right)^{2}}{\mathrm{d}s}\right\rangle \,.
\]
Therefore, combining the above results,
\[
\lim_{\mathrm{d}s\to0}\left\langle \frac{\left(\delta N\right)^{2}}{\mathrm{d}s}\right\rangle =\phi_{{\rm top}}\lim_{\mathrm{d}s\to0}\left\langle \frac{\left(\delta g\right)^{2}}{\mathrm{d}s}\right\rangle +\lim_{\mathrm{d}s\to0}\left\langle \left(1-\Theta\right)g^{2}\frac{\delta\Theta}{\mathrm{d}s}\right\rangle \,.
\]
We now use the DMFT closure in Eq. (\ref{eq:dmft-var2}) at $s$ and
at $s+{\rm d}s$ to write
\[
\left\langle \left(\delta g\right)^{2}\right\rangle =\sigma^{2}\left\langle \left(\delta N\right)^{2}\right\rangle \,.
\]
Hence we obtain, using $\lim_{\mathrm{d}s\to0}\left\langle \frac{\left(\delta N\right)^{2}}{\mathrm{d}s}\right\rangle =2\left|\hat{C}'\left(0^{+}\right)\right|$,
\begin{equation}
\left(1-\sigma^{2}\phi_{{\rm top}}\right)\lim_{\mathrm{d}s\to0}\left\langle \frac{\left(\delta N\right)^{2}}{\mathrm{d}s}\right\rangle =2\left(1-\sigma^{2}\phi_{{\rm top}}\right)|\hat{C}'\left(0^{+}\right)|=\lim_{\mathrm{d}s\to0}\left\langle \left(1-\Theta\right)g^{2}\frac{\delta\Theta}{\mathrm{d}s}\right\rangle \,.\label{eq:rel_DMFT}
\end{equation}
In the right-hand side we recover the quantity $G$ introduced in
Sec. \ref{subsec:Diversities-revisited} of the main text and the
above equation reduces to Eq. (\ref{eq:fluct_div}). Note that we
can express $G$ as 
\begin{equation}
G=\lim_{\mathrm{d}s\to0}\left\langle \left(1-\Theta\right)g^{2}\frac{\delta\Theta}{\mathrm{d}s}\right\rangle =\lim_{\mathrm{d}s\to0}\left\langle \frac{\left(1-\Theta(z(s))\right)N(s+{\rm d}s)^{2}}{\mathrm{d}s}\right\rangle \,.\label{eq:formulaG}
\end{equation}
In terms of the many-body dynamics, the above equation becomes

\[
G=\lim_{\mathrm{d}s\to0}\frac{1}{S\mathrm{d}s}\sum_{i\in\mathcal{I}(s,s+{\rm d}s)}N_{i}(s+{\rm d}s)^{2}\,,
\]
where $\mathcal{I}(s,s+{\rm d}s)$ is the subset of species experiencing
a jump from rare during the interval $[s,s+{\rm d}s]$. Following
Eq. (\ref{eq:formulaG}), the coefficient $G$ can also expressed
in terms of the steady-state distribution $\mathbb{P}[g,z]$

\[
G=\int_{0}^{+\infty}\mathrm{d}g\,\mathbb{P}[g,z=0^{-}]g^{3}\,.
\]

\subsection{Generalization for $\gamma\protect\neq0$\label{subsec:Generalize-gamma}}

\begin{figure}
\begin{centering}
\includegraphics[width=0.5\textwidth]{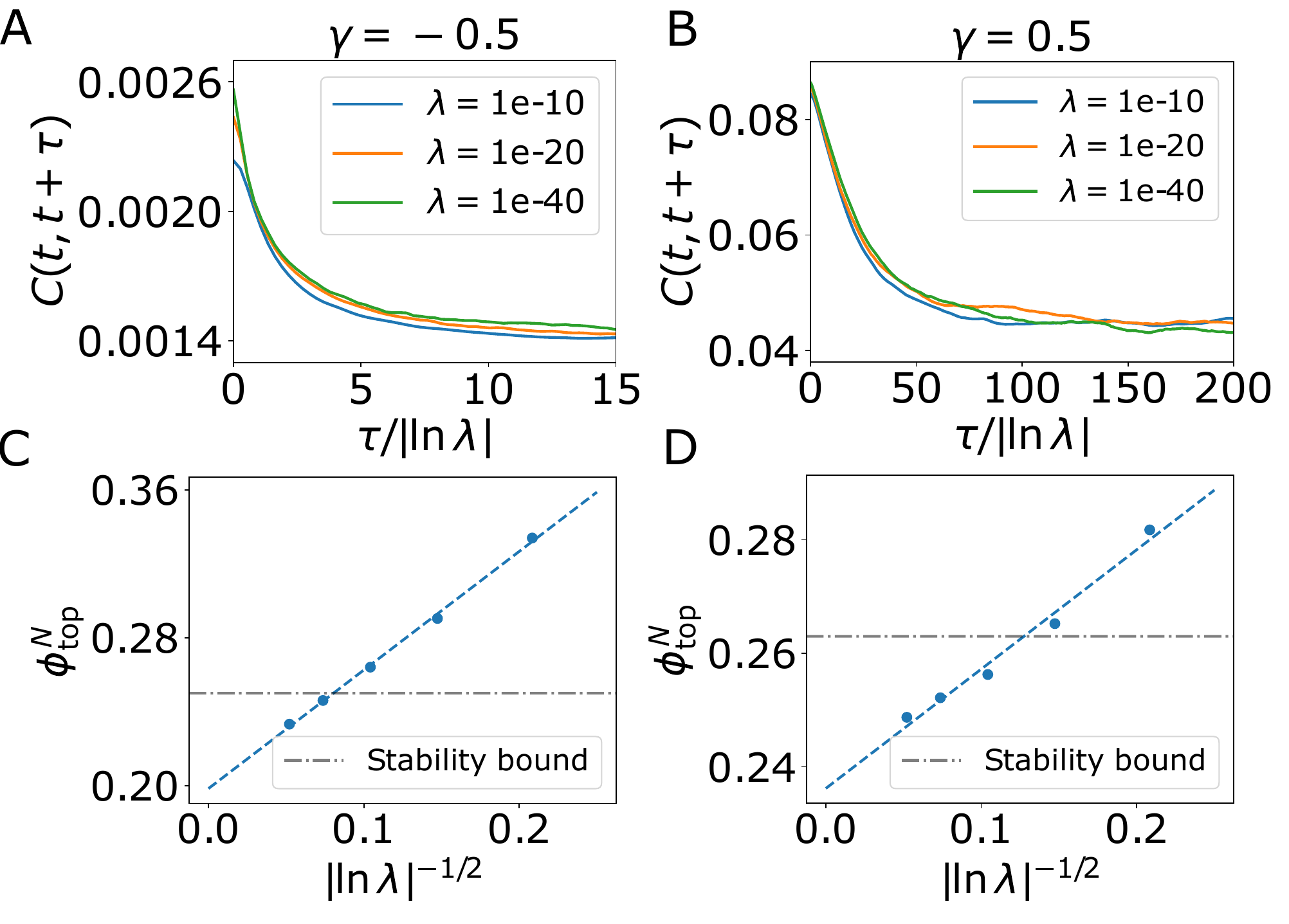}
\par\end{centering}
\caption{\label{fig:correlation_gamma}\textbf{Model behavior with partially
symmetric or antisymmetric interaction matrices.} Even when the interaction
matrix possesses some degree of symmetry or asymmetry, $\tau_{\lambda}=|\ln\lambda|$
is the only timescale controlling the dynamics when $\lambda\to0^{+}$.
We plot the collapse of the correlation function $C_{\lambda}(t,t+\tau)$
as a function of $\tau/\left|\ln\lambda\right|$ for partially anti-symmetric
($\gamma=-0.5$, A) and partially symmetric ($\gamma=0.5$, B) interaction
matrices. The top diversity, measured at finite $\lambda$, asymptotically
goes below the May bound as $\lambda\to0^{+}$ for partially anti-symmetric
($\gamma=-0.5$, C) and partially symmetric ($\gamma=0.5$, D) interaction
matrices. The diversity is evaluated by counting the number of species
with $N_{i}>10^{-3}$. The form of the extrapolation in powers of
$|\ln\lambda|$ is explained in Sec. \ref{sec:Additional-to-discussion}.
Parameters: (A) $S=5000,\,\mu=10,\,\sigma=1.3$, average over 40 realizations
(B) $S=5000,\,\mu=50,\,\sigma=4$ , average over 40 realizations.}
\end{figure}

We now consider the case $\gamma\neq0$ in the limit $\lambda\to0^{+}$.
We generalize the previous argument connecting fluctuations, diversity
and species turnover based on the many-body equations of motion and
derive Eq. (\ref{eq:rel_gamma_main-1}). We assume that the previous
scaling for the timescale of the correlation matrix holds, which we
check numerically, see Fig. \ref{fig:correlation_gamma}. In other
words, at any time $s=t/|\ln\lambda|$, the system is close to a fixed
point, meaning that some species (a fraction $\phi_{{\rm top}}$ of
them) are abundant and verify
\[
1-N_{i}(s)-\sum_{j\neq i}\alpha_{ij}N_{j}(s)=0\,,
\]
 while the others are nearly extinct, \emph{i.e. }asymptotically 
\[
N_{i}(s)=0\ .
\]
Between $s$ and $s+\mathrm{d}s$ (where $\mathrm{d}s$ is an infinitesimal
interval) we distinguish four types of populations: the ones that
were extinct at $s$ but are present at $s+\mathrm{d}s$ (that we
refer to as incoming species); the ones that are abundant at both
$s$ and $s+ds$ (that we refer to as surviving species); the ones
abundant at $s$ but rare at $s+\mathrm{d}s$ (that we refer to as
extinct species); and finally the ones rare at both $s$ and $s+ds$
that do not play any role in the following argument. At time $s$,
we have for the surviving species
\[
1-N_{i}(s)-\sum_{j\neq i}^{\mathrm{surv}}\alpha_{ij}N_{j}(s)+\sum_{j}^{\mathrm{extinct}}\alpha_{ij}N_{j}(s)=0\,,
\]
and correspondingly at time $s+\mathrm{d}s$,
\[
1-N_{i}(s+\mathrm{d}s)-\sum_{j\neq i}^{\mathrm{surv}}\alpha_{ij}N_{j}(s+\mathrm{d}s)+\sum_{j}^{\mathrm{incoming}}\alpha_{ij}N_{j}(s+\mathrm{d}s)=0\,.
\]
For all $i$ (where $i$ is a surviving species), we introduce the
notations
\[
\tilde{h}_{i}=\sum_{j}^{\mathrm{extinct}}\alpha_{ij}N_{j}(s)\,,
\]
\[
h_{i}=\sum_{j}^{\mathrm{incoming}}\alpha_{ij}N_{j}(s+\mathrm{d}s)\,.
\]
Therefore, for the surviving species
\[
\delta N_{i}\equiv N_{i}(s+\mathrm{d}s)-N_{i}(s)=(\mathrm{Id}+\boldsymbol{\alpha}^{*})_{ij}^{-1}(h_{j}-\tilde{h}_{j})\,,
\]
where $\boldsymbol{\alpha}^{*}$ is $\boldsymbol{\alpha}$ reduced
to the surviving species. We can now write the quadratic variation
of the population sizes over all species
\[
\lim_{\mathrm{d}s\to0}\frac{1}{S}\sum_{i=1}^{s}\frac{\left(\delta N_{i}\right)^{2}}{\mathrm{d}s}=\lim_{\mathrm{d}s\to0}\frac{1}{S}\sum_{i=1}^{\mathrm{surv}}\frac{\left(\delta N_{i}\right)^{2}}{\mathrm{d}s}+\lim_{\mathrm{d}s\to0}\frac{1}{S}\sum_{i=1}^{\mathrm{extinct}}\frac{N_{i}(s)^{2}}{\mathrm{d}s}+\lim_{\mathrm{d}s\to0}\frac{1}{S}\frac{1}{\mathrm{d}s}\sum_{i=1}^{\mathrm{incoming}}N_{i}(s+\mathrm{d}s)^{2}\,,
\]
where we explicitly used the fact that for the incoming species $\delta N_{i}=N_{i}(s+\mathrm{d}s)$
and that $\delta N_{i}=-N_{i}(s)$ for the extinct ones. The number
of incoming species scales as $O(\mathrm{d}s)$ and for them $N_{j}(s+\mathrm{d}s)\sim O(1)$
owing to the jump dynamics identified previously (for $\gamma=0$).
Therefore we expect the perturbing field induced by the incoming species
to scale as
\[
h_{i}\sim\sqrt{\mathrm{d}s}\,,
\]
a result in agreement with the scaling $\delta g\sim\sqrt{\mathrm{d}s}$
used in the previous section. Therefore, the species going extinct
between $s$ and $s+\mathrm{d}s$ must at time $s$ have a population
size of the order of $N_{i}(s)\sim\sqrt{\mathrm{d}s}$. As a consequence,
\[
\lim_{\mathrm{d}s\to0}\frac{1}{S}\sum_{i=1}^{\mathrm{extinct}}\frac{N_{i}(s)^{2}}{\mathrm{d}s}=0\,,
\]
because the fraction of extinct species scales as $O({\rm d}s).$
The perturbing field induced by the extinct species is thus much smaller
than the one induced by the incoming ones,
\[
\tilde{h}_{i}\sim\mathrm{d}s\,.
\]
Lastly, for the surviving species, we have
\[
\lim_{\mathrm{d}s\to0}\frac{1}{S}\sum_{i=1}^{\mathrm{surv}}\frac{\left(\delta N_{i}\right)^{2}}{\mathrm{d}s}=\lim_{\mathrm{d}s\to0}\frac{1}{S}\sum_{i=1}^{\mathrm{surv}}\sum_{j,k}^{\mathrm{surv}}(\mathrm{Id}+\boldsymbol{\alpha}^{*})_{ij}^{-1}(\mathrm{Id}+\boldsymbol{\alpha}^{*})_{ik}^{-1}\frac{h_{j}h_{k}}{\mathrm{d}s}\,.
\]
Based on the DMFT analysis of the $\gamma=0$ case, we assume that
the $O(\sqrt{\mathrm{d}s})$ perturbing fields $h_{j}$ are statistically
independent of the state of the system at time $s$ and that to leading
order in $\mathrm{d}s$, $h_{j},h_{k}$ are uncorrelated for $j\neq k$.
Therefore we obtain,
\[
\lim_{\mathrm{d}s\to0}\frac{1}{S}\sum_{i=1}^{\mathrm{surv}}\frac{\left(\delta N_{i}\right)^{2}}{\mathrm{d}s}=\lim_{\mathrm{d}s\to0}\left(\frac{1}{S}\sum_{i}\frac{h_{i}^{2}}{\mathrm{d}s}\right)\frac{\phi_{{\rm top}}}{S^{*}}\mathrm{Tr}\left[(\mathrm{Id}+\boldsymbol{\alpha}^{*})^{-1}\left((\mathrm{Id}+\boldsymbol{\alpha}^{*})^{-1}\right)^{t}\right]\,.
\]

We furthermore assume that to compute the trace we can take $\boldsymbol{\alpha}^{*}$
to be sampled from the same ensemble as the full interaction matrix
$\alpha$ (albeit with a smaller size), neglecting the correlations
induced by the dynamical selection of the community of surviving species.
Relying on the cavity method, the validity of this approximation has
been argued in the fixed point phase \citep{bunin_Ecological_2017,barbier_Fingerprints_2021}
and more generally at the level of the average spectral density, see
\citep{biroli_Marginally_2018} (the possible existence of two outlying
eigenvalues \citep{barbier_Fingerprints_2021} is sub-extensive in
the trace). We thus have,
\[
\frac{1}{S^{*}}\mathrm{Tr}\left[(\mathrm{Id}+\boldsymbol{\alpha}^{*})^{-1}\left((\mathrm{Id}+\boldsymbol{\alpha}^{*})^{-1}\right)^{t}\right]=\left[\left(\frac{1+\sqrt{1-4\gamma\phi_{{\rm top}}\sigma^{2}}}{2}\right)^{2}-\phi_{{\rm top}}\sigma^{2}\right]^{-1}\,.
\]
Furthermore,
\[
\lim_{\mathrm{d}s\to0}\left(\frac{1}{S}\sum_{i}\frac{h_{i}^{2}}{\mathrm{d}s}\right)=\lim_{\mathrm{d}s\to0}\left(\frac{\sigma^{2}}{S\mathrm{d}s}\sum_{i=1}^{\mathrm{incoming}}N_{i}(s+\mathrm{d}s)^{2}\right)\,.
\]
Together, we obtain
\begin{align*}
2\left|\hat{C}'\left(0^{+}\right)\right| & =\left(1+\phi_{{\rm top}}\sigma^{2}\left[\left(\frac{1+\sqrt{1-4\gamma\phi_{{\rm top}}\sigma^{2}}}{2}\right)^{2}-\phi_{{\rm top}}\sigma^{2}\right]^{-1}\right)\lim_{\mathrm{d}s\to0}\left(\frac{1}{S\mathrm{d}s}\sum_{i=1}^{\mathrm{incoming}}N_{i}(s+\mathrm{d}s)^{2}\right)\,,\\
 & =\frac{\left(1+\sqrt{1-4\gamma\phi_{{\rm top}}\sigma^{2}}\right)^{2}}{\left(1+\sqrt{1-4\gamma\phi_{{\rm top}}\sigma^{2}}\right)^{2}-4\phi_{{\rm top}}\sigma^{2}}\lim_{\mathrm{d}s\to0}\left(\frac{1}{S\mathrm{d}s}\sum_{i=1}^{\mathrm{incoming}}N_{i}(s+\mathrm{d}s)^{2}\right)\,.
\end{align*}
Hence, we recover Eq. (\ref{eq:rel_gamma_main-1}) of the main text
\begin{equation}
2|\hat{C}'\left(0^{+}\right)|=\frac{\left(1+\sqrt{1-4\gamma\phi_{{\rm top}}\sigma^{2}}\right)^{2}}{\left(1+\sqrt{1-4\gamma\phi_{{\rm top}}\sigma^{2}}\right)^{2}-4\phi_{{\rm top}}\sigma^{2}}G\,,\label{eq:relation_gamma}
\end{equation}
with
\[
G=\lim_{\mathrm{d}s\to0}\frac{1}{S\mathrm{d}s}\sum_{i\in\mathcal{I}(s,s+{\rm d}s)}N_{i}(s+{\rm d}s)^{2}\,,
\]
and where $\mathcal{I}(s,s+{\rm d}s)$ is the subset of species experiencing
a jump from rare during the interval $[s,s+{\rm d}s]$. For $\gamma=0$,
Eq. (\ref{eq:relation_gamma}) reduces to Eq. (\ref{eq:rel_DMFT})
derived within the DMFT formalism.Additional figures and details for
the Disscussion section\label{sec:Additional-to-discussion}

\subsection{Convergence with $S$ and $\lambda$\label{sec:mean_var}}

\begin{figure}
\begin{centering}
\includegraphics[width=0.5\textwidth]{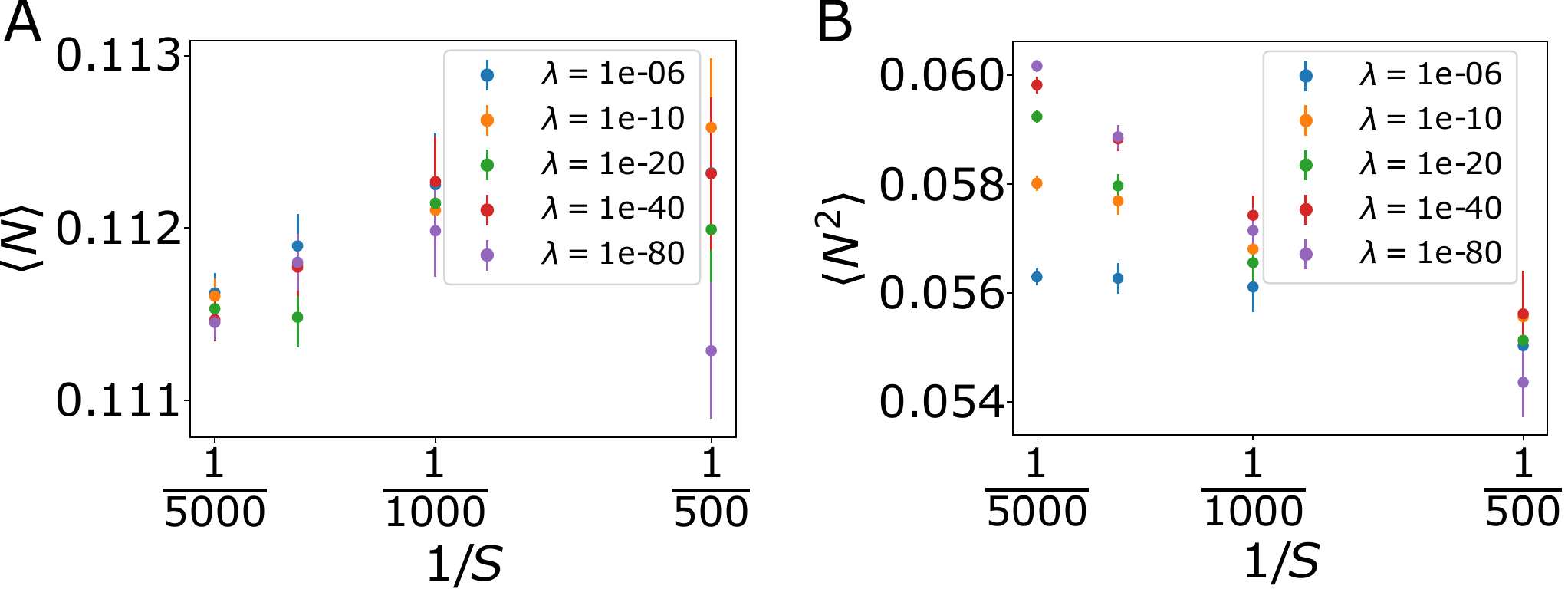}
\par\end{centering}
\caption{\label{fig:N_N2}\textbf{Behavior of the first and second moments
of the population size}. $\left\langle N\right\rangle $ and $\left\langle N^{2}\right\rangle $
are robust predictions that weakly depend on the number of species
$S$ and the migration rate $\lambda$. In the inspected range of
parameters, the variations in $\left\langle N\right\rangle $ are
of the order of 1\% and those in $\left\langle N^{2}\right\rangle $
of the order of 10\%. Parameters: $\sigma=1.8$, $\mu=10$, average
over 80 realizations.}
\end{figure}

Fig. \ref{fig:N_N2}, presents the convergence of $\left\langle N\right\rangle ,\left\langle N^{2}\right\rangle $,
showing that it is quantitatively quite robust to changes in $\lambda,S$.
One implication is that other definitions of diversity besides than
species richness, will be quite robust. For example, defining diversity
using the inverse Simpson index $S^{-1}\left(\sum_{i}N_{i}\right)^{2}/\sum_{i}N_{i}^{2}$
that is approximately $\left\langle N\right\rangle ^{2}/\left\langle N^{2}\right\rangle $,
is relatively robust in $S,\lambda$, due to the robustness of $\left\langle N\right\rangle ,\left\langle N^{2}\right\rangle $.

As briefly discussed in Sec. \ref{sec:Discussion}, to measure $\phi_{{\rm top}}$
at finite $S$ and $\lambda$, we test two options. The first is simply
a threshold on the abundance: counting all the species with $N_{i}\geq\epsilon$
at a given time, for some chosen $\epsilon\ll1$. Let $\phi_{{\rm top}}^{\text{N}}(S,\lambda,\epsilon)$
be the diversity measured this way. The second measure, denoted by
$\phi_{{\rm top}}^{\text{g}}(S,\lambda,\epsilon)$, utilizes the invasion
growth rate $g_{i}$. In it, in addition to $N_{i}\geq\epsilon$ we
also requires $g_{i}>0$. For both measures, we find a similar form
for the dependence on $\lambda,S,\epsilon$. For $\phi_{{\rm top}}^{\text{N}}$,
\begin{equation}
\phi_{{\rm top}}^{\text{N}}(S,\lambda,\epsilon)=\phi_{{\rm top}}+c_{1}\left|\ln\lambda\right|^{-\beta_{\text{N}}}+c_{2}/S+A\epsilon\,.\label{eq:eq:finite_ab_filtering-1}
\end{equation}
where $A=P(N=0^{+})$, $c_{1,2}$ are constants, and the exponent
$\beta_{\text{N}}=1/2$. The expression for $\phi_{{\rm top}}^{\text{g}}$
is of the same form, only with $\beta_{\text{g}}=1$. This form of
convergence can be understood as follows. We find that the asymptotic
distribution $P(z)$ diverges as $(-z)^{-1/2}$ when $z\to0^{-}$.
This can be traced back to the fact that $\dot{z}=0$ when $z(s)$
leaves the confining wall. Therefore, by defining a cutoff $N\geq\epsilon$,
or equivalently $z\geq\ln\epsilon/|\ln\lambda|$, the error made in
sampling the distribution $P(z)$ for $z<0$ scales as $\sqrt{|\ln\epsilon/\ln\lambda|}$,
hence the result in Eq. (\ref{eq:eq:finite_ab_filtering-1}). However,
when conditioned on $g>0$, the distribution $\mathbb{P}(g>0,z)$
is found to be finite when $z\to0^{-}$ stemming from the fact that
$\dot{z}>0$ when $z(s)$ reaches the wall from below. Thus, by defining
a cutoff $N\geq\epsilon$, or equivalently $z\geq\ln\epsilon/|\ln\lambda|$,
the error made in sampling the distribution $\mathbb{P}(g>0,z)$ for
$z<0$ only scales as $|\ln\epsilon/\ln\lambda|$, hence the exponent
$\beta_{\text{g}}=1$. These properties of the steady-state distribution
$\mathbb{P}(g,z)$ seem to be robust features of persistent random
walkers confined by hard obstacles and have been discussed in other
contexts \citep{arnoulx2023run,Wagner_2017,ezhilan_alonso-matilla_saintillan_2015}.
Since in practice $\left|\ln\lambda\right|$ would not be a very large
number, the difference in the convergence in $\lambda$ is important.
Quantitatively, $\phi_{{\rm top}}^{\text{N}},\phi_{{\rm top}}^{\text{g}}$
are typically larger than $\phi_{{\rm top}}$, and the convergence
in $\lambda$ of $\phi_{{\rm top}}^{\text{g}}$ is indeed much faster,
again highlighting the relevance of $g$. We find that the measure
$\phi_{{\rm top}}^{\text{N}}(S,\lambda,\epsilon)$ is often above
the stability bound, in contrast to the asymptotic $\phi_{{\rm top}}$
which is always below it. $\phi_{{\rm top}}^{\text{g}}(S,\lambda,\epsilon)$
can be either above or below this bound, depending on the parameters,
see Fig. \ref{fig:abundance_filtering-1-1}.
\begin{figure}
\begin{centering}
\includegraphics[width=0.5\textwidth]{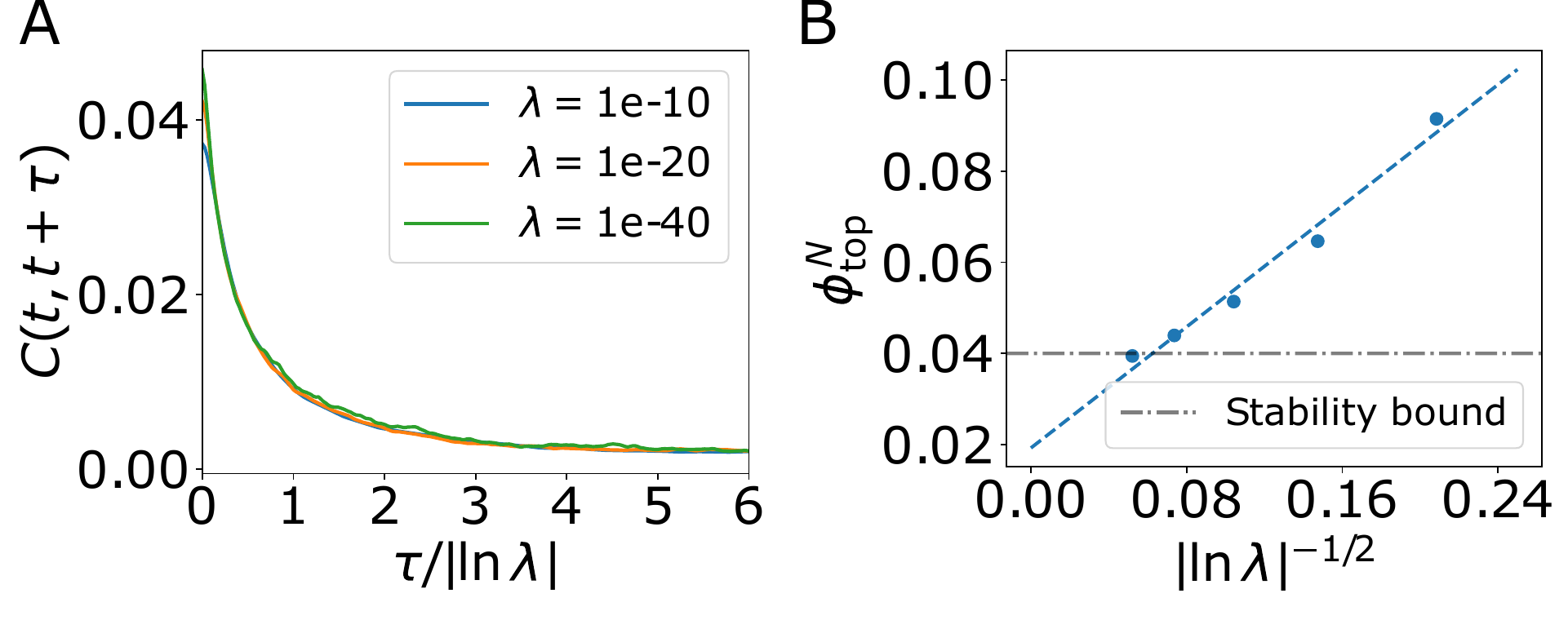}
\par\end{centering}
\caption{\label{fig:large_params}\textbf{Model behavior at higher values of
$\sigma,\mu$.} (A) Collapse of the correlation function $C_{\lambda}(t,t+\tau)$
as a function of $\tau/\left|\ln\lambda\right|$, showing that the
only timescale in the problem scales as $\left|\ln\lambda\right|$.
Compare with Fig. \ref{fig:dynamics-1}(C). (B) $\phi_{{\rm top}}^{\text{N}}$
as a function of $\lambda$, showing the same dependence on $\lambda$,
converging as $\left|\ln\lambda\right|^{-1/2}$. Compare with Fig.
\ref{fig:abundance_filtering-1-1}(B). The extrapolated asymptotic
value at $\lambda\to0^{+}$ is well-below the May bound. Here, $\sigma=5$,
$\mu=50$, $S=5000$. $\phi_{{\rm top}}^{\text{N}}$ is defined with
$N_{i}>\epsilon=10^{-3}$.}
\end{figure}

\subsection{Identifying top species by selection and subsequent dynamics\label{sec:id_top}}

The separation of top species from the rest is guaranteed for $\lambda\to0^{+}$.
Yet as discussed in Sec. \ref{sec:robustness}, for finite $S,\lambda$
this separation is not very clear when looking at the abundances,
see $P(N)$ in Fig. \ref{fig:distrib-conv}(A). By using insights
gleaned from the theory, we have shown, see Sec. \ref{sec:robustness}
and Fig. \ref{fig:abundance_filtering-1-1}, that this separation
can be much better defined even away from the asymptotic limit, by
also considering $g_{i}$, the invasion growth rate defined in Sec.
\ref{subsec:Species-abundance-and}. Here we show, that if we can
manipulate the system, by removing certain judiciously-chosen species
and rerunning the dynamics, the asymptotic distribution $P(N)$ can
be remarkably well reproduced at reasonable $S,\lambda$. Beyond potential
applications to experiments, this seems to suggest that the separation
between abundant species and the others is still present, even when
it seems blurred with other measures.

\begin{figure}
\begin{centering}
\includegraphics[width=0.5\textwidth]{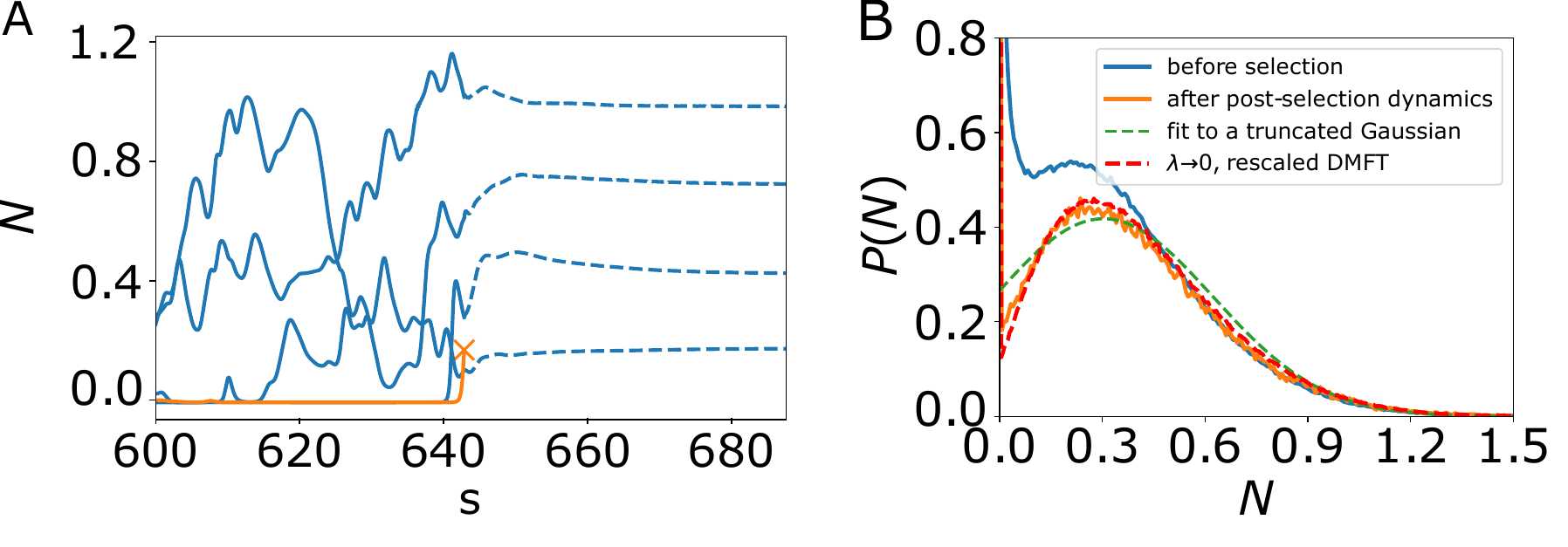}
\par\end{centering}
\caption{\textbf{Identifying the top species by combined selection and dynamics.}
(A) Top species are identified by stopping the dynamics (solid lines),
and removing species that are about to grow and disrupt the system,
here the orange trajectory, and those with a negative invasion growth
rate (by removing species $i$ if $N_{i}/g_{i}<1/2$), and then continuing
the dynamics (dashed lines). The outcome is an equilibrium. (B) The
distribution of abundances thus obtained is close to the asymptotic
distribution obtained from the rescaled dynamics. It is much closer
than the bare abundance distribution $P(N)$, and is also very different
from the distribution in the equilibrium phase. $S=5000$, $\sigma=1.8$,
$\mu=10$, $\lambda=10^{-10}$.\label{fig:Identifying-top-1}}
\end{figure}

The lack of clear separation is most pronounced at not-very-small
$N$ (say $N\sim0.1$). There, one finds: (1) abundant species, (2)
species that grow from rare and are about to disrupt the abundant
species, and (3) species with negative growth rate that are leaving
the abundant subset. The abundant species (group 1) are characterized
by appreciable $N$, and being at a fixed point, $g\simeq N$. They
occupy an equilibrium that is fully stable if not disrupted by species
that grow from rare (group 2). These can have significant $g$, with
still $N$ small compared to $g$. The idea here is to remove species
(2) and (3), which we do by removing all species that have $N_{i}/g_{i}<1/2$,
corresponding to species with positive growth rate that are in the
midst of their jump, below halfway, and those with negative growth
rate in Fig. \ref{fig:The-rescaled-dynamics.},\ref{fig:finite-migration-Trajectories}.
Then, we run again the dynamics, and we find that the remaining species
reach a stable equilibrium. The properties of the equilibrium obtained
in this way are remarkably similar to those predicted by the asymptotic
theory, even for reasonable $\lambda$. Fig. \ref{fig:Identifying-top-1}
shows the obtained abundance distribution $P(N)$. It is similar to
the asymptotic distribution, and in contrast quite different from
the instantaneous unfiltered $P(N)$ from Fig. \ref{fig:distrib-conv}(A).
Furthermore, the resulting distribution is distinct from those in
the fixed point phase (when $\sigma<\sqrt{2}$), as shown by a comparison
with the predicted distribution there, a truncated Gaussian. The asymptotic
species richness $\phi_{\text{top}}=0.28$ is recovered within 1\%;
this is lower than the stability bound, at $\phi=0.31$.

Numerical methods\label{sec:Numerical-methods}

Here we detail the numerical procedures used in producing the figures.
They are of two types:

(1) Full many-variable simulations of Eq. (\ref{eq:multibody}).

(2) Numerical solution of the rescaled dynamics, defined in full in
Eqs. (\ref{eq:DMFT2},\ref{eq:dmft-mean2},\ref{eq:dmft-var2}).

\medskip{}

(1) We used the explicit Runge-Kutta method of order 5(4) implemented
by the ODE solver scipy.solve\_ivp in Python, to simulate Eq. (\ref{eq:multibody}).

(2) The set of equations of the rescaled dynamics are self-consistent:
the trajectory $z(s)$ depends on $g(s)$, which is sampled with correlation
function $\hat{C}(s,s')$ and mean $m(s)$. Self-consistently, $\hat{C}(s,s'),\,m(s)$
depend on the statistics of $z(s)$, see Eqs. (\ref{eq:dmft-mean2},\ref{eq:dmft-var2}).
This self-consistency is standard in DMFT formulations. We used a
well-known numerical method to solve it \citep{eissfeller_New_1992,roy_Numerical_2019}.
It starts with a guess for $\left\langle g(s)g(s')\right\rangle ,\,\left\langle g(s)\right\rangle ,$
generates realizations of $g(s)$ and from that trajectories $z(s)$,
which are then used to update $\left\langle g(s)g(s')\right\rangle ,\,\left\langle g(s)\right\rangle $.
This is repeated until convergence.

In practice, the DMFT simulations were carried with: a timestep ${\rm d}s=0.5$
for $s<500$, ${\rm d}s=0.1$ for $500\leq s<600$ and ${\rm d}s=0.05$
for $600\leq s<700$. We used (i) 500 iterations with averaging over
1000 realizations and injection fraction $0.3$ followed by (ii) 1000
iterations with averaging over 10000 realizations and injection fraction
$0.3$ followed by (iii) 1500 iterations with averaging over 10000
realizations and injection fraction $0.03$ and followed by (iv) 2000
iterations with averaging over 10000 realizations and injection fraction
$0.003$. In order to precisely obtain the diversity graph in Fig.
\ref{fig:richness-SAD-1}(G) we initialized the algorithm with $\left\langle g(s)g(s')\right\rangle ,\,\left\langle g(s)\right\rangle $
given by their fixed point branch value with $\left\langle g(s)g(s')\right\rangle $
destabilized by a small identity matrix (corresponding to small amplitude
white noise). The results agree very well with the asymptotic values
found from full simulations of Eq. (\ref{eq:multibody}), when $S\to\infty,\lambda\to0^{+}$
are taken carefully, see Fig. \ref{fig:abundance_filtering-1-1}.

\medskip{}

\emph{Simulation details for individual figures} of the main text:

Fig. \ref{fig:richness-SAD-1}. (A) $S=1000$, $\sigma=1$, $\mu=10$,
$\lambda=10^{-8}$ (B) $S=1000$, $\sigma=2$, $\mu=10$, $\lambda=10^{-8}$
(C, E) $S=20000$, $\sigma=1$, $\mu=10$, average over 200 realizations.
(D, F) $S=20000$, $\sigma=2$, $\mu=10$, average over 200 realizations.
(G) DMFT numerics of the rescaled equation at $\mu=10$.

Fig. \ref{fig:dynamics-1}: (A) $S=5000$, $\lambda=10^{-10}$, $\sigma=2$,
$\mu=10$. (B) $S=20000$, $\lambda=10^{-10}$, $\sigma=2$, $\mu=10$,
average over 40 realizations (C) $S=20000$, $\lambda=0$, $\sigma=2$,
$\mu=10$, average over 40 realizations. (D) $S=20000$, $\sigma=2$,
$\mu=10$, average over 40 realizations. DMFT line obtained with the
$\lambda\to0^{+}$ rescaled dynamics.

Fig. \ref{fig:distrib-conv}: same as in Fig. \ref{fig:dynamics-1}.
(A,C) $S=20000$, $\lambda=0$, $\sigma=2$, $\mu=10$, average over
40 realizations. DMFT line obtained with the $\lambda=0$ rescaled
dynamics. (B, D) $S=20000$, $\sigma=2$, $\mu=10$, average over
40 realizations. DMFT line obtained with the $\lambda\to0^{+}$ rescaled
dynamics.

Fig. \ref{fig:crossover}: Same as in Fig. \ref{fig:dynamics-1}.
S=20000, $\sigma=2$, $\mu=10$, average over 40 realizations. All
species start at $N_{i}(t=0)=0.5$.

Fig. \ref{fig:abundance_filtering-1-1}: $\sigma=1.8$, $\mu=10$,
average over 40 realizations.

Fig. \ref{fig:finite-migration-Trajectories}: $S=5000$, $\sigma=1.8$,
$\mu=10$, $\lambda=10^{-6}$.

\bibliographystyle{unsrt}
\bibliography{refs}

\end{document}